# Selectively pulsed spin order transfer increases parahydrogen-induced NMR amplification of insensitive nuclei and makes polarization transfer more robust

Andreas B. Schmidt,[1,2,3‡]* Arne Brahms,[4‡] Frowin Ellermann,[3] Stephan Berner,[1] Jürgen Hennig,[1] Dominik von Elverfeldt,[1] Rainer Herges,[4] Jan-Bernd Hövener,[3] Andrey N. Pravdivtsev[3]*

1 Department of Radiology, Medical Physics, Medical Center, University of Freiburg, Faculty of Medicine, University of Freiburg, Killianstr. 5a, Freiburg 79106, Germany.

2 German Cancer Consortium (DKTK), partner site Freiburg and German Cancer Research Center (DKFZ), Im Neuenheimer Feld 280, Heidelberg 69120, Germany.

3. Section Biomedical Imaging, Molecular Imaging North Competence Center (MOIN CC), Department of Radiology and Neuroradiology, University Medical Center Kiel, Kiel University, Am Botanischen Garten 14, 24118, Kiel, Germany.

4. Otto Diels Institute for Organic Chemistry, Kiel University, Otto-Hahn-Platz 5, 24118, Kiel, Germany.

**ABSTRACT:** We describe a new method for pulsed spin order transfer (SOT) of parahydrogen induced polarization (PHIP) that enables close to 100 % polarization in incompletely $^{2}$H-labeled molecules by exciting only the desired protons in a frequency-selective manner. While a selective pulse (SP) on $^{1}$H at the beginning of pulsed SOT had been considered before, using SPs during the SOT suppresses undesired indirect spin-spin interactions. As a result, we achieved a more robust SOT for the SP variants of the phINEPT+ sequence that we refer to as phSPINEPT+. Thereby, for the first time, we report a sequence that is effective for all weakly coupled spin systems. Our simulations show that the method converts close to 100 % of the parahydrogen-derived spin order into $^{13}$C hyperpolarization in weakly coupled three-spin systems and partially or fully $^{2}$H-labeled molecules if relaxation is neglected. Experimentally we demonstrate high hyperpolarization of $^{13}$C with 15.8 % for 1-$^{13}$C-hydroxyethyl propionate-d$_3$ and 12.6 % for 1-$^{13}$C-ethyl acetate-d$_6$, which corresponds to ≈47 % and ≈38 % if the enrichment of parahydrogen had been 100 %. Even in non-$^{2}$H-labeled molecules, a remarkable $^{13}$C polarization is achieved, e.g. up to 20 % were simulated for 100 % pH$_2$, and 1.25 % were obtained experimentally for 1-$^{13}$C-ethyl pyruvate and 50 % pH$_2$, which can be further improved by a faster hydrogenation. As a result, full deuterium labeling may no longer be required e.g., when new PHIP agents are investigated, the synthesis of fully deuterated molecules is too complex, or when a kinetic isotope effect regarding the metabolic conversion rate of an agent is to be avoided. Using SPs during SOT seems very promising and may be extended to other sequences in the context of PHIP and beyond to make them less prone to experimental imperfections or real molecular environments.

Nuclear magnetic resonance (NMR) has become indispensable in chemistry and medical diagnostics (magnetic resonance imaging; MRI), although it effectively detects only a millionth of the present nuclear spins. The intrinsically low thermal polarization of nuclear spins, which is of the order of parts per million, often limits magnetic resonance to detect only the most abundant substances and nuclei – for MRI, typically hydrogen atoms in water and lipids. Hence, the quest for an efficient hyperpolarization and spin order transfer (SOT) has been ongoing for several decades. Hyperpolarization has already enabled several exciting applications e.g., in analytical[1] and catalytic chemistry,[2,3] clinical in vivo gas imaging,[4] as well as observation of metabolism in real time, early-stage cancer diagnostics, and longitudinal monitoring of treatment and recovery.[5–7] Among current techniques, the versatile parahydrogen induced polarization (PHIP),[8] also referred to as parahydrogen and synthesis allows a dramatically enhanced nuclear alignment (PASADENA),[9] and signal amplification by reversible exchange (SABRE)[10] are very promising, because they are cost and time efficient.[2,11] PHIP and SABRE provided the basis for liquid state polarization of more than 50 % for $^{13}$C and over 20 % for $^{15}$N.[12–14] Both techniques utilize the highly ordered spin alignment of parahydrogen (pH$_2$), i.e. of molecular hydrogen (H$_2$) enriched in the singlet nuclear spin isomer. PHIP incorporates the spin order by a pairwise addition of pH$_2$ to an unsaturated compound (with C=C or C≡C bond), which results in an $^{1}$H hyperpolarized product (**Fig. 1a**). SABRE exploits reversible chemical exchange and interaction of pH$_2$ and the target molecule.[10] Hence, SABRE does not change the chemical structure of the compound and has enabled continuous hyperpolarization.[15–19] However, for MR-applications, a heteronuclear polarization transfer e.g., to $^{13}$C or $^{15}$N is often needed, because (a) the lifetime of their polarization is typically longer than for $^{1}$H, (b) the background signal of non-hyperpolarized tissue *in vivo* is negligible, and (c) monitoring of metabolic conversion is easier because the chemical shift spread is larger. To transfer the spin order of pH$_2$ to X-nuclei (nuclei other than $^{1}$H), two approaches are commonly used: "spontaneous" transfer by evolution at specific external magnetic fields – i.e. magnetic field cycling (MFC)[20] – or specific manipulations of the spin system by RF pulses.[13,21–32] The latter features some interesting advantages, which are: (a) hyperpolarized agents can be prepared directly at high field in the MR system and thus, no external "polarizer" is needed;[33] (b) high-quality, commercially available hardware can be used to conduct the SOT (i.e. NMR spectrometers or MRI systems); (c) a typically higher theoretical polarization (i.e., if the molecule consists of more than the three coupling nuclear spins from $pH_2$ and $^{13}C$ in MFC polarization transfer is often limited, e.g. $P_{13C}$ = 30 % for ethyl acetate);[34] (d) fully deuterium ($^2H$) labeled precursors can be polarized effectively with the advantage that the lifetime of hyperpolarization is often longer, which is favorable for (bio-) medical monitoring of hyperpolarized agents *in vivo*.[35] Despite the nuclear spin-1 of the $^2H$ atom the $^2H$-$^1H$ and $^2H$-$^{13}C$ J-coupling interactions are effectively suppressed with high-field pulsed SOT because no $^2H$ refocusing pulses are applied.[23] Although, deuteration is advantages for the pulsed polarization transfer it can cause a kinetic isotope effect (e.g. altered metabolic rates or changes of the pharmokinetics).[36,37] However, in current high-field pulsed methods polarization transfer can suffer, if more $^1H$ than the two stemming from $pH_2$ are present. This is because undesired $^1H$-$^1H$ J-coupling interactions were effective throughout the SOT(see SI). Here, we propose and demonstrate a method that partially solves this problem physically instead of chemically with $^2H$-labeling:

$pH_2$ and the use of Selective Pulses enabled effective and robust Insensitive Nuclei Enhancement by Polarization Transfer (phSPINEPT+, in accordance to the predecessor – the nonselective phINEPT+ sequence; [21] **Fig. 1b**). Compared to the phINEPT+ sequence, the use of frequency-selective excitation and refocusing pulses holds fundamental benefits. First, the maximum polarization yield is more than doubled, allowing up to ~100 % $^{13}C$-polarization for fully or partially $^2H$-labeled molecules (**Fig. 2c,d**). Experimentally, $^{13}C$-polarizations up to ~16 % were achieved, corresponding to $P_{13C}$ = 47 % if 100 % parahydrogen had been used. Second, even when no $^2H$ labeling is present, a robust and remarkably high polarization is possible e.g., close to 20% for 1-$^{13}C$-ethyl pyruvate (SI, Fig. S6). Here, the theoretical maximum is reduced from 100 % to 20 % only, because the selective pulses also excite the $^1H$ of the original vinyl-CH or vinyl-$CH_2$ group and J-couplings are not fully suppressed. Thus, selectively pulsed SOT improves the polarization for (partially) protonated molecules, e.g. for the research and investigation of new PHIP agents for which isotope labeling is synthetically difficult, too expensive[38] or undesirable.[36,37] Additionally, for such molecules our phSPINEPT+ variants are more robust to uncertainties or experimental errors compared to phINEPT+ (**Fig. 2c,d**) and other non-selective SOTs (i.e. the evolution timings $\tau_1$ and $\tau_2$ are a function of all J-coupling constants between excited nuclei of the molecule; see SI).

Known SOT sequences for PASADENA (and SABRE) are usually designed for strongly coupled (i.e. where proton-proton indirect spin coupling is much larger than the chemical shift difference between protons $H_I$ and $H_S$, $J_{H_IH_S} \gg \Delta\delta$) or weakly coupled spin systems ($J_{H_IH_S} \ll \Delta\delta$). All of the latter are modifications of the INEPT NMR sequence,[39] which account for the very different spin order at the beginning of the sequence. For a thermally polarized spin only a single-spin order ($I_z$) is present; for two hydrogens I and S of $pH_2$ after hydrogenation – if the added $^1H$ are weakly coupled – two-spin order $I_zS_z$ is given.[40] Among the sequences for weakly coupled spin systems, phSPINEPT+ and phINEPT+ sequences carry the advantage of consisting only of two free evolution intervals ($2\tau_1$ and $2\tau_2$, **Fig. 1**). In contrast, the two other currently known SOT sequences for weakly coupled spin systems, namely the efficient spin order transfer to heteronuclei via relayed INEPT chains (ESOTHERIC) and the selective excitation of polarization using PASADENA (SEPP) -INEPT sequence need three intervals (SI, Fig. S4) and 1-2 more RF-pulses. Note that the concept of exploiting a first frequency-selective pulse to excite the $I_zS_z$ spin order has indeed been introduced with SEPP.[22,41,42] For high-field SABRE selective X-nuclear pulses have been used to re-polarize and accumulate polarization by repeating SABRE-INEPT.[43,44] However, the selective $^1H$ refocusing and excitation during the SOT of $pH_2$ spin order is a new concept that enabled above-mentioned advantages of selective polarization transfer.

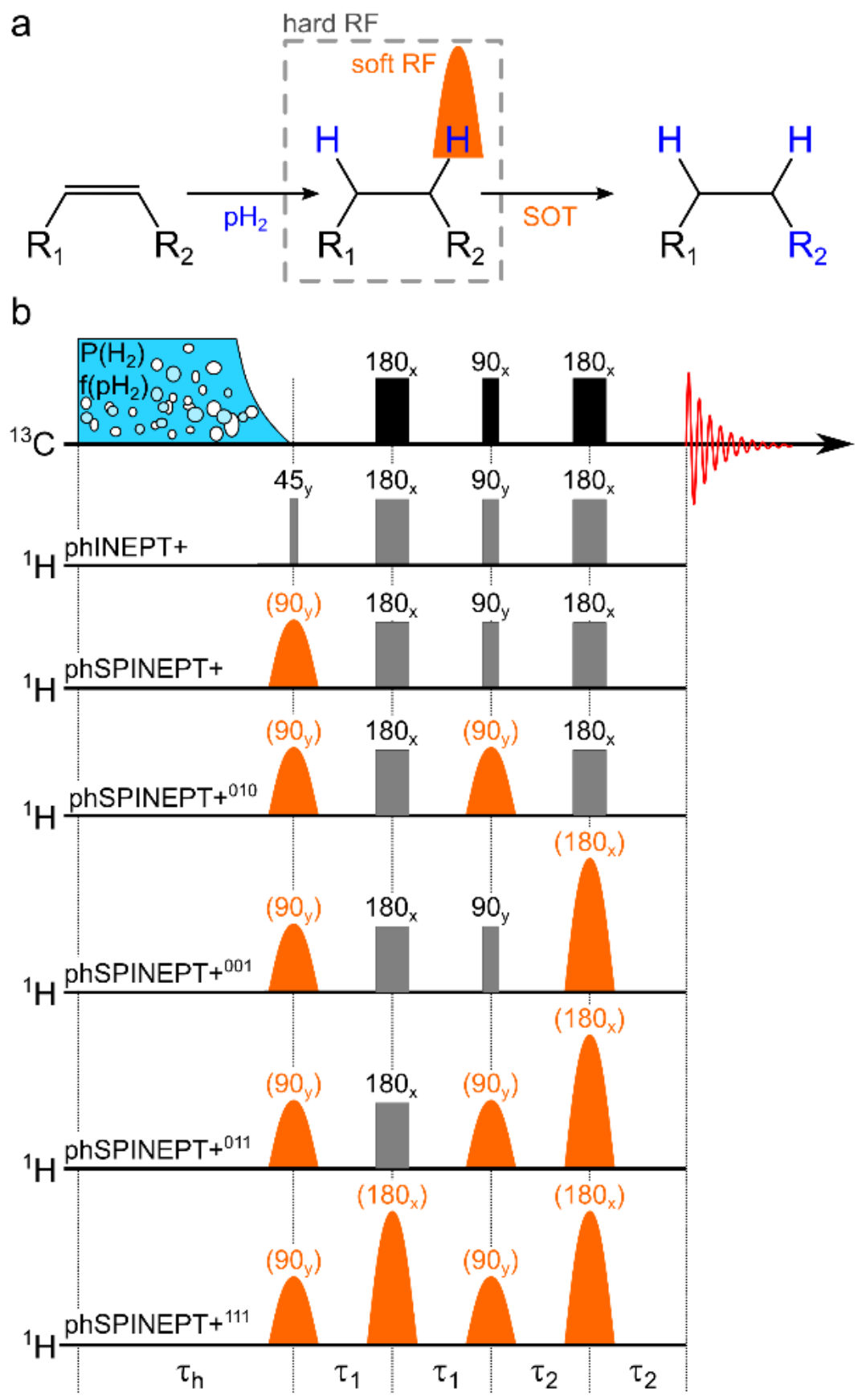

**FIGURE 1. Hydrogenation and spin order transfer (SOT) scheme (a) and phINEPT+ and phSPINEPT+ sequences (b).** After hydrogenation with $pH_2$, spin order of hyperpolarized spins is transferred to the selected $^{13}C$ nuclei using an SOT sequence of choice (polarized sites are indicated in blue). phINEPT+ and several variants of phSPINEPT+$^{xyz}$ sequences are shown, which all consist of radiofrequency pulses (with flip angles 45, 90, or 180° and phases x or y) resonant on $^1H$ or $^{13}C$ and time intervals of free

evolution ($\tau_1$, $\tau_2$; $\tau_h$ is the time during which the precursor is hydrogenated). Note that the shaped orange pulses excite only one of the two $pH_2$-nascent $^1H$ selectively and are usually 100 times longer than other, rectangular "hard" pulses. At the end of the SOT sequences enhanced $^{13}C$ signal is detected; alternatively, $^{13}C$ polarization could be flipped at this time along the static magnetic field direction to use the hyperpolarized molecules for MR imaging applications.

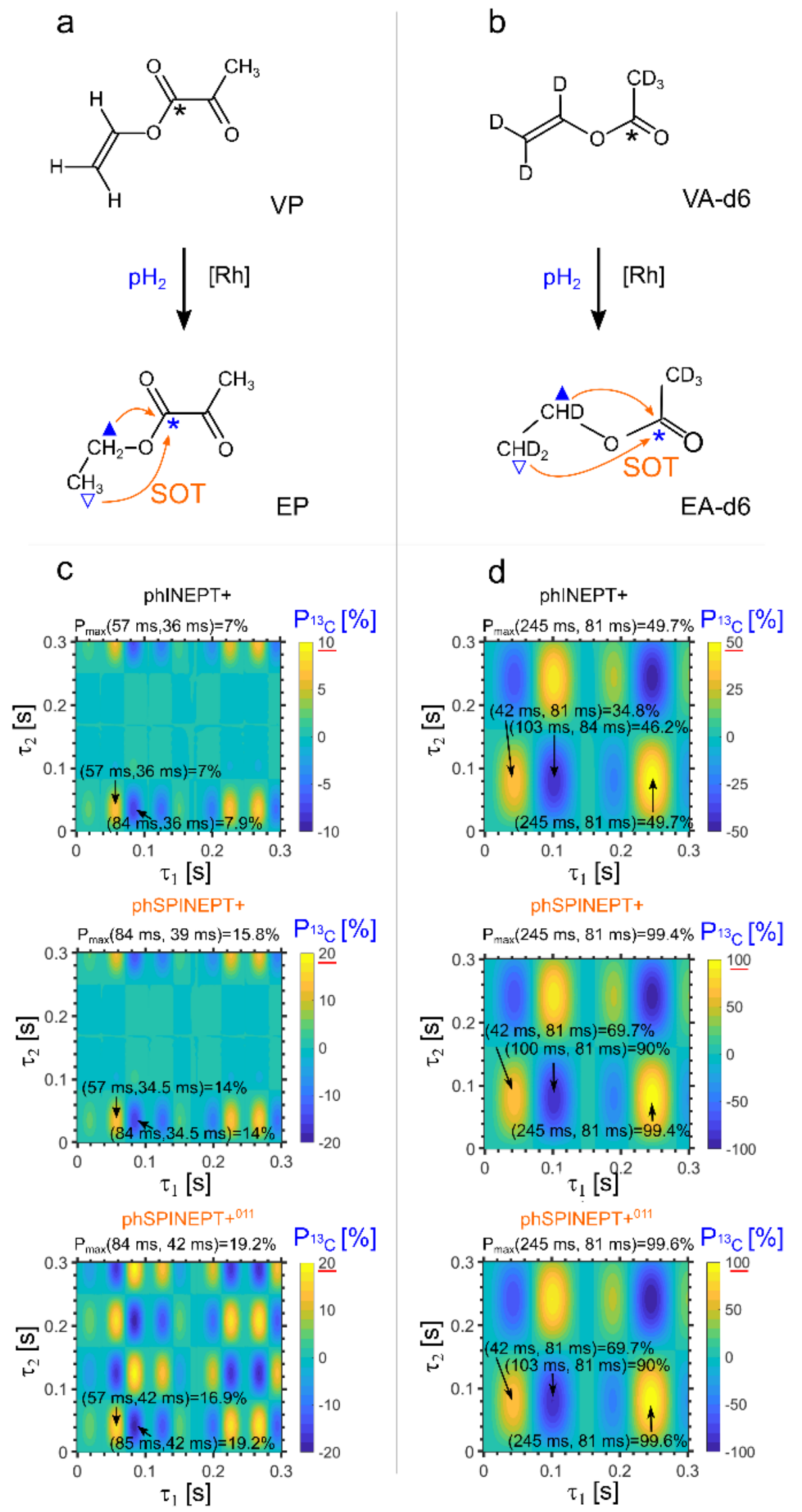

**FIGURE 2. Reaction schemes and simulated $^{13}C$ polarization yield.** Hydrogenation of vinyl pyruvate (VP, a) and vinyl acetate-$d_6$ (VA-$d_6$, b) and simulated performance of phINEPT+ and phSPINEPT+ sequences for 1-$^{13}C$-ethyl pyruvate (EA, c) and ethyl acetate-$d_6$ (1-$^{13}C$-EA, d). Selective pulses increase the $^{13}C$ polarization by more than two times compared to conventional phINEPT+. In addition, some J-coupling effects are "decoupled" using the phSPINEPT+ sequences, which makes the sequences less sensitive to evolution timings $\tau_1$ and $\tau_2$ (NMR parameters, simulation details and simulations for other PHIP molecules and SOT sequences are provided in the SI).

phSPINEPT+ is achieved in two steps: first, the hard 45° excitation in the beginning of phINEPT+ sequence is exchanged by a selective 90° pulse resonant on only one of the $pH_2$-nascent $^1H$ nuclei.[42] Second, other $^1H$-pulses of the sequence can be played out selectively as well (**Fig. 1b**). The effect of using one or both of the last two $^1H$-pulses in a frequency-selective way ensures that polarization during the $\tau_2$-period is only transferred between the excited $^1H$ and the $^{13}C$ (phSPINEPT+$^{001/010/011}$ **Fig. 1b**). All other J-coupling interactions are completely refocused and do not disturb the polarization transfer. As a result, the polarization transfer is more stable and effective, which is clearly seen in a broader and more even polarization transfer in the $\tau_2$-dimension (**Fig. 2c**). Our simulations showed that the first refocusing $^1H$ pulse of the phSPINEPT+ sequences must excite both $pH_2$ protons. This concept ensures that J-coupling evolution converts the two-spin order originating from $pH_2$ into magnetization of the one of the two $^1H$ that was selectively excited. Consequently, the phSPINEPT+$^{111}$, in which all $^1H$ pulses are selective, provided close to zero transfer of net polarization to the desired $^{13}C$ (SI, Fig. S5,6,7,9). Note that the first refocusing $^1H$ pulse could, in principle, excite only the two $pH_2$ protons selectively to suppress J-couplings with ambient non-$pH_2$-nascent $^1H$ further, but this concept was not further explored here.

Naturally, an even more robust transfer was achieved when fully $^2H$-labeled molecules were considered: for all sequences oscillations in $\tau_1$ and $\tau_2$-dimensions featured a much lower frequency e.g., for vinyl acetate-$d_6$ (VA-$d_6$) hydrogenated to ethyl acetate-$d_6$ (EA-$d_6$, **Fig. 2d**) compared to vinyl pyruvate (VP) hydrogenated to ethyl pyruvate (EP, **Fig. 2c**). In addition to prolonged lifetime of hyperpolarization and a higher maximum polarization yield, this more-robust polarization transfer is another advantage of using (partially) deuterated agents with high-yield pulsed SOT sequences.

To prove our findings experimentally, we explored the polarization transfer following the reactions of VP $\xrightarrow{pH_2}$ EP, VA-$d_6$ $\xrightarrow{pH_2}$ EA-$d_6$ and 1-$^{13}C$-hydroxyethyl acrylate-$d_3$ hydrogenated to 1-$^{13}C$-hydroxyethyl propionate-$d_3$ (HEA-$d_3$$\xrightarrow{pH_2}$HEP-$d_3$, **Fig. 3**). As predicted by the theory, in all our experiments with phSPINEPT+ the signal enhancement was more than two times higher than with phINEPT+; the corresponding polarization values were 1.25 % vs 0.44 % for VP, 12.6 % vs 5.7 % for VA-$d_6$ and 15.8 % vs 6.4 % for HEP-$d_3$. All cases the sequences were optimized to transfer polarization to the carboxyl carbon (1-$^{13}C$ position), which has the longest longitudinal relaxation time $T_1$ and therefore is advantageous for applications in MRI. Note that VP and VA-$d_6$ were not $^{13}C$ enriched.

It is noticeable that our experimental polarizations are lower than suggested from our simulations. Partially, this effect was caused by the limited experimental enrichment of $pH_2$ that featured only 50%.[11] If the enrichment had been 100%, which was assumed in the simulations and is possible with dedicated hardware,[45–47] three times higher polarization values would have been achieved.[48] Another important source of discrepancy is that relaxation was neglected in our simulation. The relaxation of $pH_2$, i.e. para-to-ortho conversion, in catalyst solution before the hydrogenation is negligible on the time scale of seconds.[49] However, relaxation after the addition of $pH_2$ and before spin-order is transferred needs to be considered. The polarization at time $t$ after hydrogenation can be estimated to[33]

$$P(\mathrm{t}) \cong \exp[-R(I_Z S_Z)t] \cdot P_{\max}$$

Here $R(I_Z S_Z)$ is the relaxation rate of $pH_2$ spin order that is used for polarization transfer, in our case it is $I_Z S_Z$ PASADENA spin order and $P_{max}$ is a theoretically maximum polarization of $^{13}C$ that can be produced by SOT (detailed in SI, section 5). $R(I_Z S_Z)$ is of the order of $T_1$ of two protons. The exact value depends on the present relaxation mechanisms [50,51] and needs to be measured experimentally. Here we simply estimate it to be an average $T_1$ of the two added protons. Hence, for the used hydrogenation time of 8 s we estimated $P(t)$ for HEP-$d_3$, EA-$d_6$, and EP using $R(I_Z S_Z)$ =0,086 $s^{-1}$, 0.076 $s^{-1}$, and 0.22 $s^{-1}$, respectively (SI, Fig. S1). By setting $P_{max}$ to values predicted by our simulations for the reported phSPINEPT+ experiments in Fig. 3 (SI, Table S1), the model suggests $P(t{=}8\ \mathrm{s}) \approx 44$ %, 48 %, and 3 % for HEP-$d_3$, EA-$d_6$, and EP, respectively – 47.4 %, 37.8 %, and 3.75 % would have been achieved experimentally if 100 % $pH_2$ had been used. Hence, relaxation of $^1H$ during the hydrogenation likely accounts for most of the discrepancy of our simulated and experimental polarization. As a consequence, this result clearly demonstrates the utmost importance of a fast hydrogenation using a dedicated high-pressure and high-temperature reactor[12,33] along with molecules, isotope labeling and reaction conditions that feature long proton relaxation times.

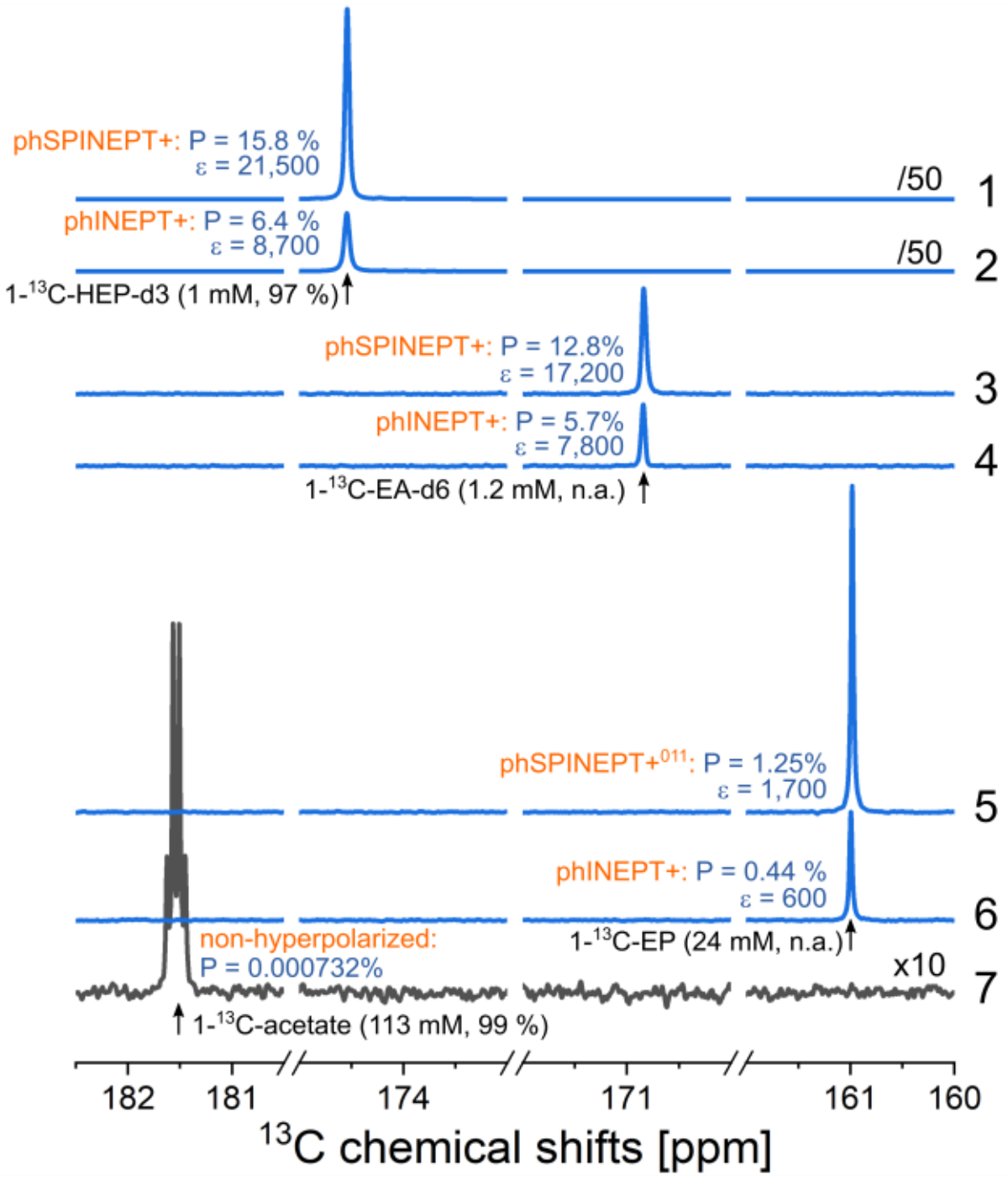


**FIGURE 3. Hyperpolarized and reference $^{13}C$ NMR spectra.** phINEPT+ and phSPINEPT+ spectra of $^{13}C$ labeled 1-$^{13}C$-hydroxyethyl propionate-$d_4$ (1-$^{13}C$-HEP-$d_3$, 1 and 2), $^{13}C$ natural abundance ethyl acetate-$d_6$ (1-$^{13}C$-EA-$d_6$, 3 and 4) , $^{13}C$ natural abundance ethyl pyruvate (1-$^{13}C$-EP, 5 and 6) and reference 1-$^{13}C$-acetate spectrum (7). The concentrations, $^{13}C$ enrichment or natural abundance (n.a.), polarization values, signal enhancement at 9.4 T and type of the SOT sequence are indicated. Experimental details and parameters are given in **SI**, **Tab S1**.

Another exciting advantage of phSPINEPT+ is that it enables close to 100 % polarization in weakly coupled spin systems for both, completely asymmetrically (e.g. EA-$d_6$, with J-couplings between $^{13}C$ and the $^1H$ stemming from $pH_2$ $|J^3_{HC}| \gg |J^4_{HC}|$) and molecules with a relatively symmetric spin-spin coupling pattern (HEA-$d_3$, $|J^3_{HC}| \cong |J^4_{HC}|$; SI, Fig. S1). Such a versatility was not reached by the previously known sequences (SI, Fig. S8,10,12).[34] Thus, phSPINEPT+ appears to be the long-sought sequence (or general method) that can be applied to achieve maximum output for all weakly coupled PHIP molecules, including metabolic agents like esters of pyruvate, acetate and lactate[52–54], as well as agents for angiography (HEP)[35] or functional receptor imaging.[55] As the method only requires two input parameters (i.e. the $\tau_1$ and $\tau_2$ evolution timings), finding optimal parameters by means of quantum-mechanical simulation in the 9-spin systems to obtain ideal experimental performance is straight-forward (e.g. using our MOIN-spin library[56] and the dedicated phINEPT+ and phSPINEPT+ scripts, SI). However, our simulations suggest that in some cases maximum $^{13}C$ polarization can be achieved with a slightly shorter SOT sequence using SEPP-INEPT+ or ESOTHERIC (SI, Fig. S10). Ultimately, the ideal SOT scheme is always a trade-off between duration of SOT, its theoretical efficiency and real, experimental performance. Therefore, we suggest comparing the performance of the sequences for each molecule of interest theoretically followed by experimental testing and fine tuning. While comparing the sequences for molecules with different coupling patterns will be useful, it is beyond the scope of this article and will be part of our future investigations.

We foresee that other pulse sequences – in the context of PHIP and SABRE but also in other fields were RF sequences are used – would as well benefit from selective excitations and refocusing. Potentially using more free evolution intervals in the SOT sequences the robustness of polarization transfer can be further improved. Here, however, it shall be noted that the SPINEPT concept works only if it is possible to excite one and not both of the $pH_2$-originating $^1H$. This may limit the application to higher magnetic fields and molecules with a sufficiently large chemical shift difference Δδ. Here, we tested the sequences for HEP-$d_3$ at 9.4 T (Δδ=1.2 ppm, corresponding to ~480 Hz; for all other investigated molecules, Δδ is larger). If lower fields or molecules with smaller Δδ were explored, long or differently shaped pulses would possibly be needed to achieve a narrower excitation of only one $^1H$. At the same time, selective excitation may even turn out as a significant advantage i.e., for biomedical PHIP implementations where polarization takes place in the MRI system. Here, long pulses are often used because of the limited RF power available and the poor filling factor of the RF coil.[57] As only the first refocusing RF pulse needs to simultaneously excite both $pH_2$ protons in the SPINEPT sequences, the problem of limited excitation bandwidth of clinical MRI units is tremendously reduced.

In conclusion, using frequency-selective pulses in the phINEPT+ sequence bears several advantages, namely: (i) doubling the polarization to a theoretical maximum of ~100 % in three-spin systems,[42] (ii) tolerance to parameter settings as undesired $^1H$-$^1H$ and $^1H$-$^{13}C$-couplings are suppressed, (iii) applicability to a broader range of coupling regimes, (iv) few(er) pulses thus, robust to flip angle imperfections. Disadvantages include that only weakly coupled spin systems are suitable as frequency selective excitation is necessary and that still one non-selective $^1H$ hard pulse is needed. Although, $^2H$-labeling of

nuclear sites in proximity to the $pH_2$ protons is required to maximize polarization (longer $^1H$ relaxation times; selective excitation possible), selectively pulsed SOT increased the efficacy of polarization transfer in protonated and partially deuterated compounds. For example, polarization of HEP-$d_3$ (Fig. 3) resulted in 15.8 % $^{13}C$ polarization (or 47.4 % if corrected to 100 % $pH_2$) although only 3 out of 7 protons of the precursor are deuterated.

As such, selectively pulsed SOT significantly improves the exploration and polarization of new PHIP agents. It may constitute an important step towards making the benefits of hyperpolarized MRI accessible to a wider community and clinical applications; - monitoring metabolic processes and improving diagnostics of pathologies, including cancer, at a much lower cost than before.

## ASSOCIATED CONTENT

**Supporting Information (pdf)**. Chemicals, methods (including used parameters of the SOT with simulated and experimental performance, simulation-based optimization of SOT parameters for more molecules, as well as the implemented pulse programs), and additional hyperpolarized NMR spectra. The MOIN spin library and scripts to simulate SOT performance (.zip). This material is available free of charge via the Internet at http://pubs.acs.org."

## AUTHOR INFORMATION

### Corresponding Authors

*ABS: andreas.schmidt@uniklinik-freiburg.de and
ANP: andrey.pravdivtsev@rad.uni-kiel.de

### Author Contributions

ABS, ANP: conceptualization, methodology, writing – original draft, preparation, funding acquisition. ABS, AB, FE, SB, JBH, ANP: investigation. AB conceptualization of VP synthesis. JH, DvE, RH and JBH: supervision, funding acquisition. JBH: provided the environment. All authors contributed to discussions and helped interpreting the results and have given approval to the final version of the manuscript.
‡These authors contributed equally.

## ACKNOWLEDGMENT

We acknowledge funding from German Federal Ministry of Education and Research (BMBF) within the framework of the e:Med research and funding concept (01ZX1915C), German Cancer Consortium (DKTK), DFG (SCHM 3694/1-1, HO-4602/2-2, HO-4602/3, GRK2154-2019, EXC2167, FOR5042, SFB1479, TRR287), Kiel University and the Faculty of Medicine, Research Commission of the University Medical Center Freiburg (SCHM2146-20). MOIN CC was founded by a grant from the European Regional Development Fund (ERDF) and the Zukunftsprogramm Wirtschaft of Schleswig-Holstein (Project no. 122-09-053). We thank Stefan Glöggler for providing us with VA-$d_6$ and Mariya Pravdivtseva for designing the table of content and cover image.

# Supporting materials for

# Selectively pulsed spin order transfer increases parahydrogen-induced NMR amplification of insensitive nuclei and makes polarization transfer more robust

Andreas B. Schmidt,[1,2,3‡]* Arne Brahms,[4‡] Frowin Ellermann,[3] Jürgen Hennig,[1] Dominik von Elverfeldt,[1] Rainer Herges,[4] Jan-Bernd Hövener,[3] Andrey N. Pravdivtsev[3]*

1 Department of Radiology, Medical Physics, Medical Center, University of Freiburg, Faculty of Medicine, University of Freiburg, Killianstr. 5a, Freiburg 79106, Germany.

2 German Cancer Consortium (DKTK), partner site Freiburg and German Cancer Research Center (DKFZ), Im Neuenheimer Feld 280, Heidelberg 69120, Germany.

3. Section Biomedical Imaging, Molecular Imaging North Competence Center (MOIN CC), Department of Radiology and Neuroradiology, University Medical Center Kiel, Kiel University, Am Botanischen Garten 14, 24118, Kiel, Germany.

4. Otto Diels Institute for Organic Chemistry, Kiel University, Otto-Hahn-Platz 5, 24118, Kiel, Germany.

# Contents

## 1. Chemicals

The following compounds were used without any additional purification: the substrate precursors hydroxyethyl acrylate-1-$^{13}C$,2,3,3-$d_3$ (HEA-$d_3$, CAS 1216933-17-3, Sigma Aldrich, USA), vinyl acetate (VA, CAS 108-05-4, Sigma Aldrich, USA), vinyl acetate-$d_6$ (VA-d6, CAS 189765-98-8, CDN isotopes, CA; distributed by EQ Laboratories GmbH, DE); the solvents chloroform-d (CAS 865-49-6, Sigma Aldrich, USA) and acetone-$d_6$ (CAS 666-52-4, Deutero GMBH, DE); and the hydrogenation catalyst [1,4-Bis-(diphenylphosphino)-butan]-(1,5-cyclooctadien)-rhodium(I)-tetrafluoroborat ([Rh], CAS 79255-71-3, Sigma Aldrich, USA). The substrate precursor vinyl pyruvate (VP, 80%) was synthesized in house; the protocol and quality assurance will be published elsewhere. The concentrations of the precursors and catalyst used in experiments are reported in **Tab S1** together with the achieved polarization values.

In the hyperpolarization experiments, VA and VA-$d_6$ were hydrogenated into ethyl acetate (EA) and ethylacetate-$d_6$ (EA-$d_6$), respectively; VP was reacted into ethyl pyruvate (EP); HEA-$d_3$ formed hydroxyethyl propionate-$d_3$ (HEP-$d_3$).

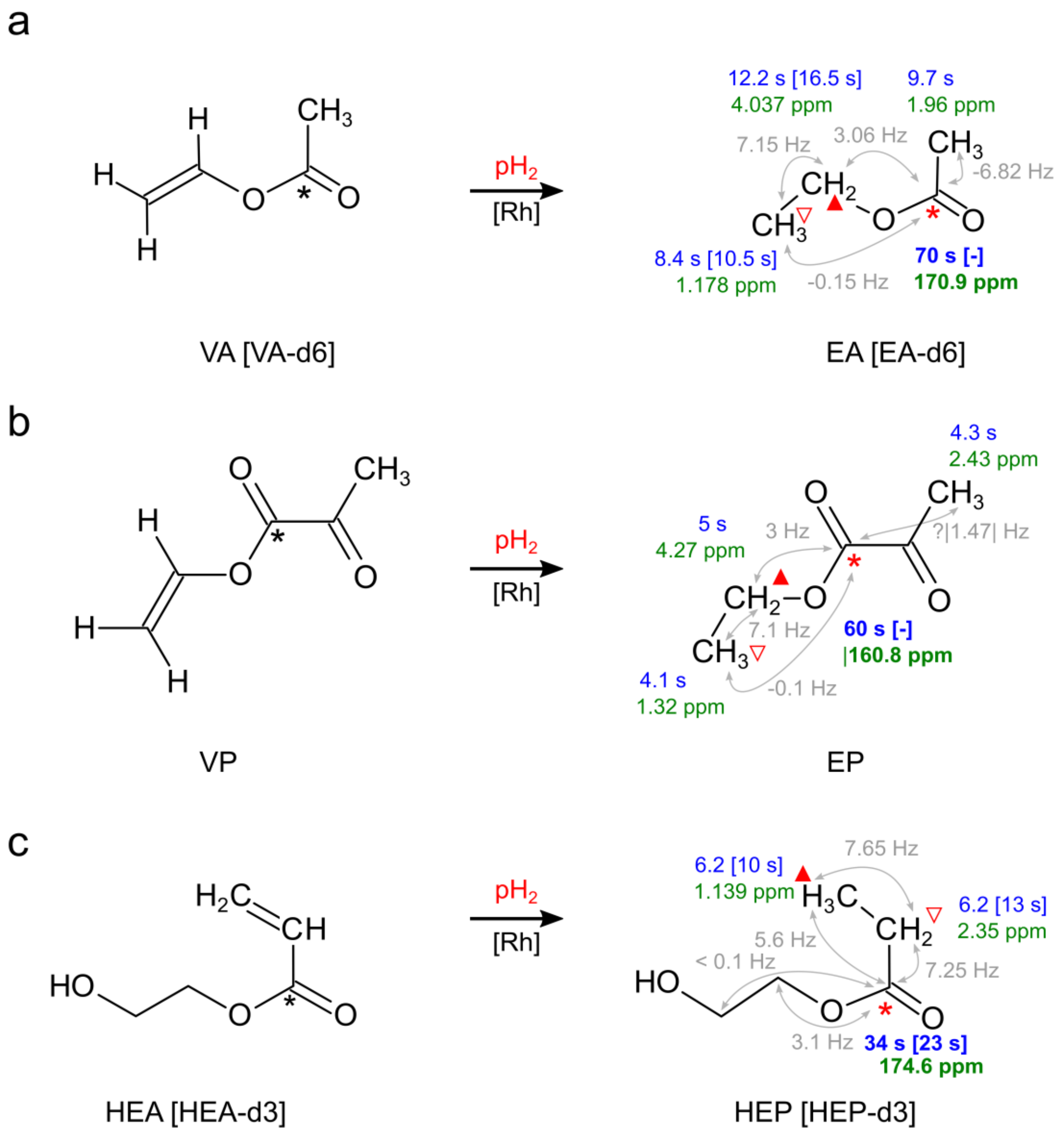


**Figure S1. Chemical structure and hydrogenation reactions of VA→EA, VP→EP, HEA→HEP. T1 relaxation times, J-coupling constants, and chemical shifts of the relevant proton and carbon-13 sites are shown (blue, gray, and green, respectively).** The J-coupling constants and chemical shifts were used to calculate polarization transfer in phINEPT+ and phSPINEPT+ experiments. Values in brackets corresponds to deuterated substrates. All $T_1$-values, chemical shifts and J-coupling constants were measured customly at 330 K in chloroform-d at 9.4 T (see **Fig S2**, fitted 1-$^{13}C$-HEP spectrum,). Only HEP-d3 $^{1}H$ $T_1$ values are reported for $^{13}C$ labeled molecules, all the rest were measured at natural abundance. The red solid and hollow triangles indicate the $pH_2$-nascent $^{1}H$; the red asterisk indicates the $^{13}C$ site of interest towards which polarization was transfered in our experiments.

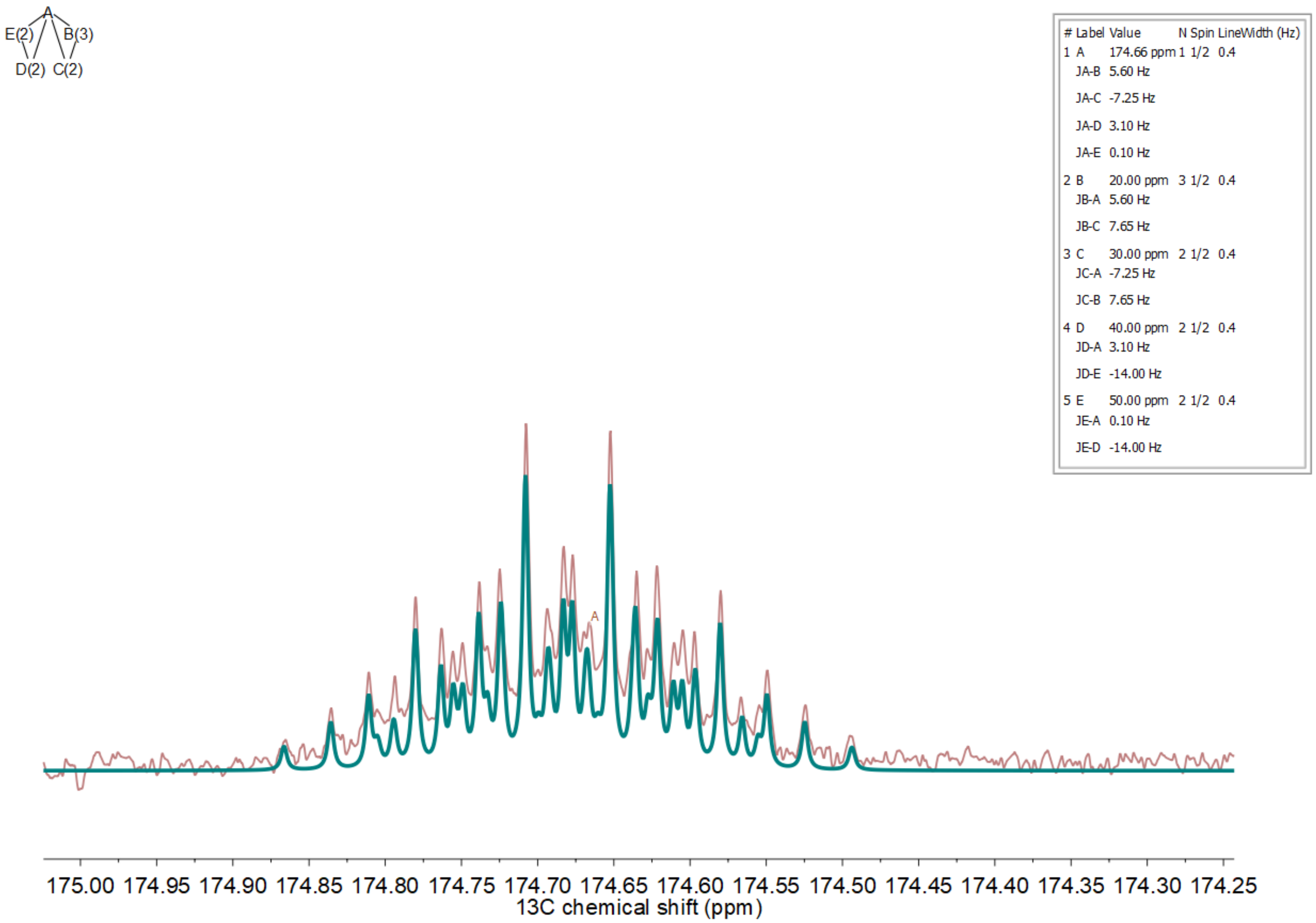


**Figure S2. $^{13}C$ NMR spectrum of 1-$^{13}C$-HEP (red, natural abundance) in chloroform-d at 330 K and 9.4 T and fitted spectrum (green).** The determined J-coupling constants are given in the insert table and in **Fig S1** and were used for the simulations.

### 2. $pH_2$ polarization setup

All experiments were carried out on a 400 MHz widebore nuclear magnetic resonance (NMR) system (WB NMR 400 MHz, Avance Neo, 9.4 T, Bruker, DE) using a 5 mm BBFO probe. A medium wall high-pressure 5 mm NMR tubes were used (524-PV-7, Wilmad-LabGlass, USA). The hyperpolarization setup (**Fig S3**) allowed flushing gaseous $pH_2$ through the samples at elevated pressure (7.8 bar). The stepwise experimental procedurewas as follows:

1. 450 μL of the catalyst-precursor containing sample were filled into the NMR tube..
2. Gas from the tubing that enters the liquid sample was replaced by purging $N_2$ for approximately 5s.
3. The NMR tube was connected to the gas system.
4. The NMR tube was positioned in the NMR spectrometer.
5. The temperature of the probe and sample were equilibrated (about 2-3 minutes, 330 K).
6. The homogeneity of the static magnetic field was optimized using an automated routine (Topshim, Topspin 4.0.9, Bruker, DE).
7. The pressure was build up in the system without flushing $pH_2$ through the sample (solenoid valves S3 and S4 were opened for 5 seconds).
8. Pressure was build up and equilibrated in the gas-inlet and -outlet path from the NMR tube (S2 and S3 open for 25 ms, subsequently all S1-S4 valves were simultaneously; the 25 ms delay procedure preventing gas from streaming into the solution)
9. $pH_2$ gas was guided through the solution (S1 and S2 open for 6 seconds).
10. Stop the gas flow by equilibrating the inlet and outlet pressure (S1-S4 open for 1-2 s).
11. Polarization transfer and signal detection was performed (S1-S4 were kept open; $^1H$ decoupling was turned on when $^{13}C$ signals were acquired).

Note that the steps 3-11 were automated and fully controlled via the pulse program and TTL outputs of the NMR system and a microcontroller-based power relais box.

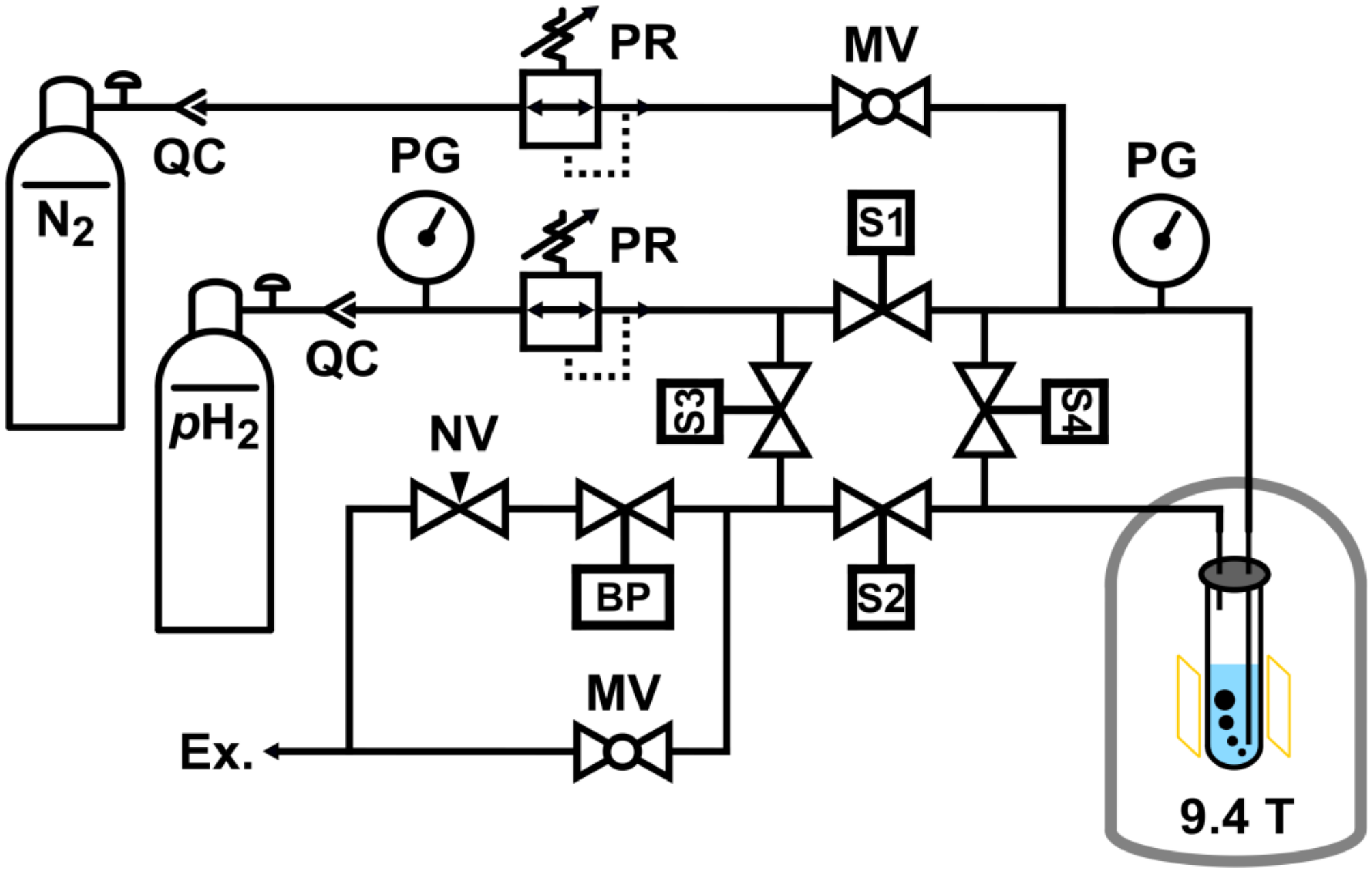


**Figure S3. Scheme of the $pH_2$ bubbling system.** The main components of the system are pressure gauges (PG), pressure regulators (PR), solenoid valves ([S]), quick connectors (QC), backpressure valve ([BP]), needle valves (NV), and manual shut-off valves (V). A part of the system can be flushed with $N_2$.

## 3. Schematics of SEPP-INEPT+ and ESOTHERIC pulse sequences

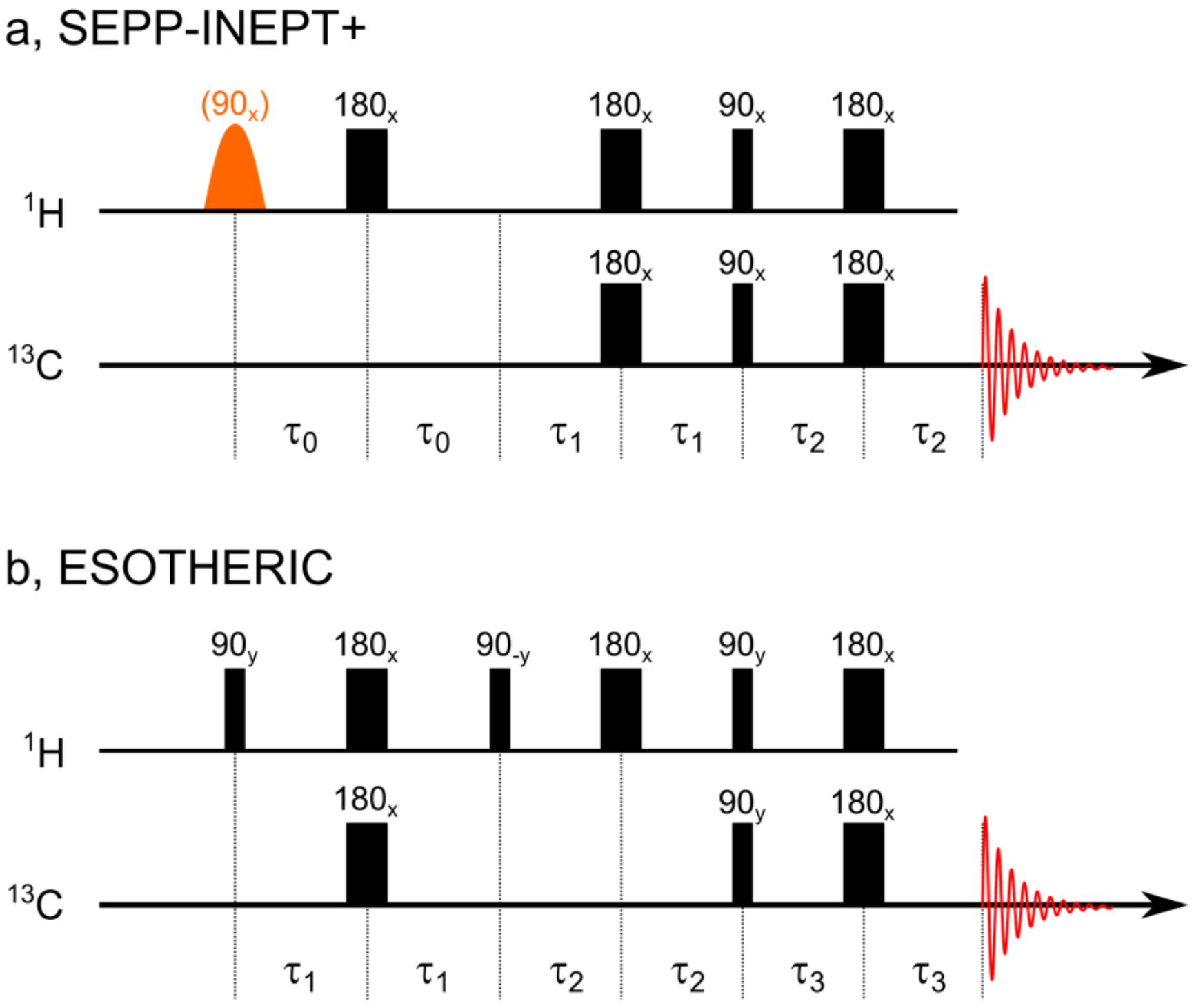


F

**Figure S4. Schematics of the SEPP-INEPT+[2] (a) and ESOTHERIC[3] sequence (b).** The sequences consist of non-selective radiofrequency hard pulses (black bars) with flip angle of 90° or 180° and phase x, y, -x, or -y on the $^{1}H$ and $^{13}C$ channel. SEPP-INEPT+ starts with a frequency selective $^{1}H$ pulse (orange).

## 4. phSPINEPT+ polarization transfer

### 4.1. Polarization transfer results

**Table S1. Experimental parameters and polarization yield for each investigated substrate and sequence**. $\boldsymbol{P^{th}}$ is the theoretical maximum $^{13}$C-polarization at perfect conditions with 100% p$H_2$; $\boldsymbol{P^{exp}}$ is the experimentally observed $^{13}$C-polarization with 50% p$H_2$ (values reported in the main text are indicated in **bold green**). If 100 % p$H_2$ would have been used (e.g. when produced with a helium cooled system[4]), the observed polarizations were 3 times larger. The same applies for the reported $^{1}$H multiplet polarization values $\boldsymbol{P_m^{th}}$, $\boldsymbol{P_m^{exp}}$. The total hydrogenation time $t_h$ corresponds to the sum of steps 9 and 10 in section 2. All experiments were conducted at 330 K, 7.8 bar p$H_2$ and 9.4 T. For HEP, some experiments were not performed (indicated with $\boldsymbol{P^{exp}} = -$ ), and the experimentally used parameters ($\tau^{exp}$) were different from the optimal values predicted by our simulations ($\tau^{th}$).

| PHIP system | PASADENA | phINEPT+ | phSPINEPT+ | phSPINEPT+$^{01}_{0}$ | phSPINEPT+$^{00}_{1}$ | phSPINEPT+$^{01}_{1}$ |
|---|---|---|---|---|---|---|
| **VA-d6→EA-d6 in $CDCl_3$**<br>c=1 mM ($^{13}$C n.a.)<br>$t_h$ = 8 s<br>$\delta_{CDH}$ = 4.1 ppm | $P_m^{th}$ = 100 %<br>$P_m^{exp}$ = 21.9 % | $\tau_1$ = 103 ms<br>$\tau_2$ = 78 ms<br>$P^{th}$ = 45 %<br>**$P^{exp}$ = 5.7 %** | $\tau_1$ = 100 ms<br>$\tau_2$ = 81 ms<br>$P^{th}$ = 90%<br>**$P^{exp}$ = 12.8 %** | $\tau_1$ = 103 ms<br>$\tau_2$ = 81 ms<br>$P^{th}$ = 90 %<br>$P^{exp}$ = 10.8 % | $\tau_1$ = 103 ms<br>$\tau_2$ = 81 ms<br>$P^{th}$ = 90 %<br>$P^{exp}$ = 12.2 % | $\tau_1$ = 100 ms<br>$\tau_2$ = 81 ms<br>$P^{th}$ = 90 %<br>$P^{exp}$ = 12.1 % |
| **1-$^{13}$C-HEA-d3→1-$^{13}$C-HEP-d3**<br><br>**in acetone-d6, unless stated otherwise**<br><br>c=1 mM (99 % $^{13}$C)<br>$t_h$ = 8 s<br>$\delta_{CDH}$ = 2.35 ppm<br>$\delta_{CD2H}$ = 1.15 ppm | $P_m^{th}$ = 100 %<br>$P_m^{exp}$ = 19.8 % | $\tau_1^{th}$ = 34 ms<br>$\tau_2^{th}$ = 19 ms<br>$P^{th}$ = 49.6 %<br><br>$\tau_1^{exp}$ = 34 ms<br>$\tau_2^{exp}$ = 17 ms<br>$P^{exp}$ = 4.8 %<br>**($P^{th}$ = 42.5 %)**<br>**in $CDCl_3$:**<br>**$P^{exp}$ = 6.4 %** | s-CDH<br>$\tau_1^{th}$ = 34 ms<br>$\tau_2^{th}$ = 172 ms<br>$P^{th}$ = 93.4%<br><br>$P^{exp}$ = – | s-CDH<br>$\tau_1^{th}$ = 34 ms<br>$\tau_2^{th}$ = 172 ms<br>$P^{th}$ = 93.4 %<br><br>$P^{exp}$ = | s-CDH<br>$\tau_1^{th}$ = 33 ms<br>$\tau_2^{th}$ = 34 ms<br>$P^{th}$ = 99.8 %<br><br>$\tau_1^{exp}$ = 36 ms<br>$\tau_2^{exp}$ = 44 ms<br>$P^{exp}$ = 10.0 %<br>($P^{th}$ = 88 %) | s-CDH<br>$\tau_1^{th}$ = 33 ms<br>$\tau_2^{th}$ = 34 ms<br>$P^{th}$ = 99.8 %<br><br>$\tau_1^{exp}$ = 36 ms<br>$\tau_2^{exp}$ = 44 ms<br>$P^{exp}$ = 9.4 %<br>($P^{th}$ = 88 %) |
| | | | s-CD2H<br>$\tau_1^{th}$ = 37 ms<br>$\tau_2^{th}$ = 141 ms<br>$P^{th}$ = 77.4 %<br><br>$\tau_1^{exp}$ = 34 ms<br>$\tau_2^{exp}$ = 98 ms<br>$P^{exp}$ = 0.3 %<br>($P^{th}$ = 0.8%) | s-CD2H<br>$\tau_1$ = 37 ms<br>$\tau_2$ = 141 ms<br>$P^{th}$ = 77.2 %<br><br>$P^{exp}$ = – | s-CD2H<br>$\tau_1^{th}$ = 38 ms<br>$\tau_2^{th}$ = 45 ms<br>$P^{th}$ = 94.7 %<br><br>$\tau_1^{exp}$ = 34 ms<br>$\tau_2^{exp}$ = 36 ms<br>$P^{exp}$ = 9.6 %<br>($P^{th}$ = 89 %) | s-CD2H<br>$\tau_1$ = 38 ms<br>$\tau_2$ = 45 ms<br>$P^{th}$ = 94.7 %<br><br>$\tau_1^{exp}$ = 34 ms<br>$\tau_2^{exp}$ = 36 ms<br>$P^{exp}$ = 9.9 %<br>**($P^{th}$ = 89 %)**<br>**in $CDCl_3$:**<br>**$P^{exp}$ = 15.8 %** |
| **VP→EP in $CDCl_3$**<br>c=30 mM ($^{13}$C n.a.)<br>$t_h$ = 8 s<br>$\delta_{CDH}$ = 4.33 ppm | $P_m^{th}$ = 100 %<br>$P_m^{exp}$ = 16 % | $\tau_1$ = 57 ms<br>$\tau_2$ = 36 ms<br>$P^{th}$ = 7 %<br>**$P^{exp}$ = 0.44 %** | $\tau_1$ = 57 ms<br>$\tau_2$ = 34.5 ms<br>$P^{th}$ = 14%<br>$P^{exp}$ = 0.95 % | $\tau_1$ = 57 ms<br>$\tau_2$ = 36 ms<br>$P^{th}$ = 13.9 %<br>$P^{exp}$ = 1.1 % | $\tau_1$ = 57 ms<br>$\tau_2$ = 42 ms<br>$P^{th}$ = 16.9 %<br>$P^{exp}$ = 1.2 % | $\tau_1$ = 57 ms<br>$\tau_2$ = 42 ms<br>$P^{th}$ = 16.9 %<br>$P^{exp}$ = 1.2 % |
| | | | | | | $\tau_1$ = 57 ms<br>$\tau_2$ = 36.5 ms<br>$P^{th}$ = 16.5 %<br>**$P^{exp}$ = 1.3 %** |
| **VA→EA in $CDCl_3$**<br>**c=30 mM ($^{13}$C n.a.)**<br>$t_h$ = 23 s – for INEPT and 8 s for PASADENA<br>$\delta_{CDH}$ = 4.15 ppm | $P_m^{th}$ = 100 %<br>**$P_m^{exp}$ = 17 %** | $\tau_1$ = 83 ms<br>$\tau_2$ = 216 ms<br>$P^{th}$ =7.7%<br>**$P^{exp}$ = 0.10 %** | $\tau_1$ = 83 ms<br>$\tau_2$ = 216 ms<br>$P^{th}$ = 15.5 %<br>**$P^{exp}$ = 0.21 %** | $\tau_1$ = 85 ms<br>$\tau_2$ = 216 ms<br>$P^{th}$ = 14.5 %<br>**$P^{exp}$ = 0.24 %** | $\tau_1$ = 85 ms<br>$\tau_2$ = 204 ms<br>$P^{th}$ = 16.5 %<br>**$P^{exp}$ = 0.31 %** | $\tau_1$ = 85 ms<br>$\tau_2$ = 204 ms<br>$P^{th}$ = 16.5 %<br>**$P^{exp}$ = 0.24 %** |

### 4.2. Polarization transfer maps (Simulations)

#### A. Methods

To simulate the polarization transfer maps listed below, we used the MOIN-spin library[5]. All necessary J-coupling constants and chemical shifts are given in **Fig. S1.** RF pulses with infinitesimally small durations were used. No relaxation effect or magnetic field inhomogeneity were considered. 100% $pH_2$ was assumed. The initial state of the system was calculated at $B_0$ by removing all coherences in the Eigen basis of liquid state Hamiltonian of the system.

#### B. 1-$^{13}$C-Vinyl acetate to 1-$^{13}$C-ethyl acetate

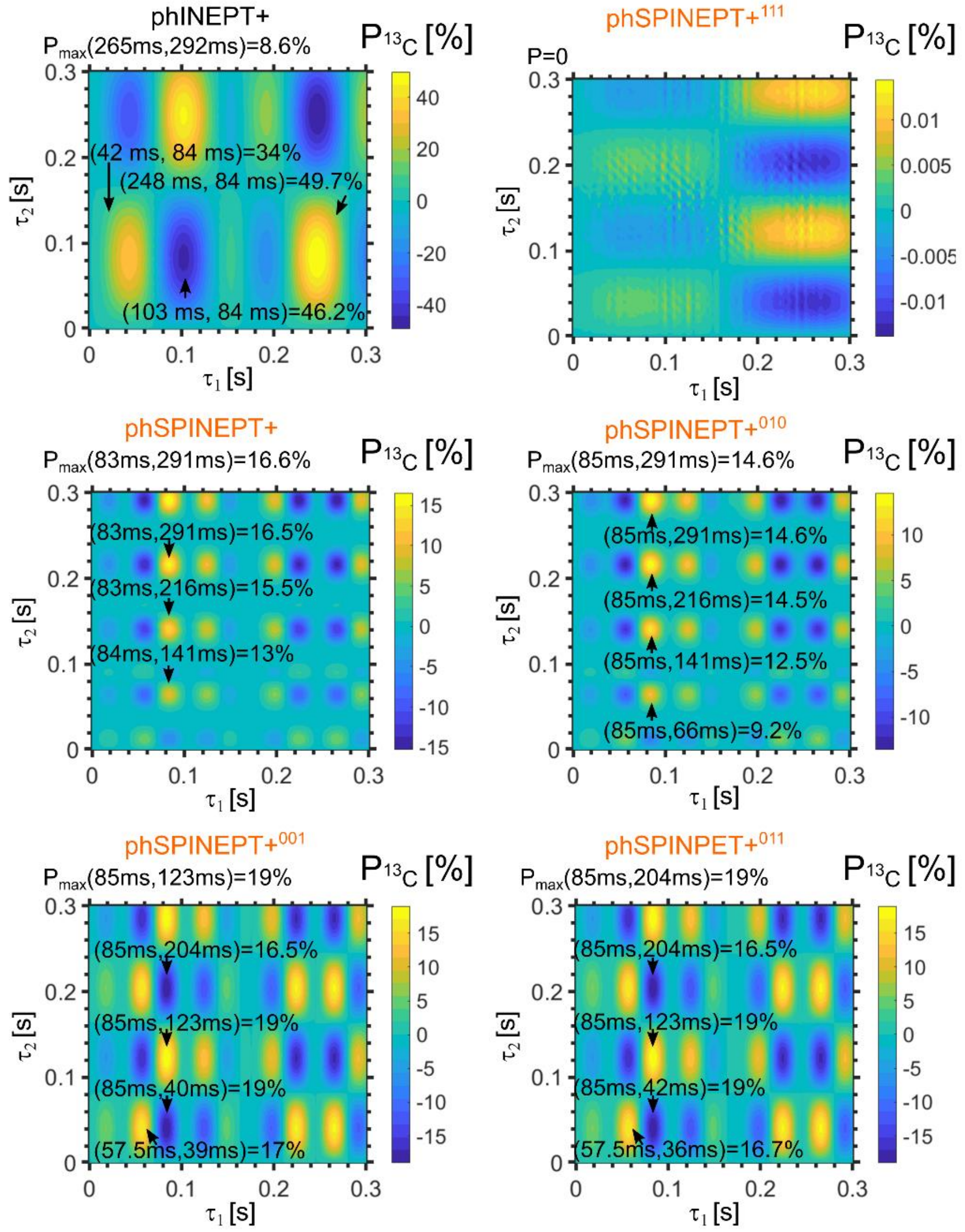


**Figure S5. $^{13}$C-polarization yield of 1-$^{13}$C-EA using phINEPT+ or phSPINEPT+ as a function of the two evolution time intervals $\tau_1$ and $\tau_2$.** Eight protons (two stemming from $pH_2$) and one carbon at the acetate C1 position were considered (**Fig. S1**). The selective radio frequency pulses were set in resonance with the $CH_2$ protons of EA at the chemical shift of $\delta$ = 4.15 ppm.

### C. 1-$^{13}$C-Vinyl acetate-$d_6$ to 1-$^{13}$C-ethyl acetate-$d_6$

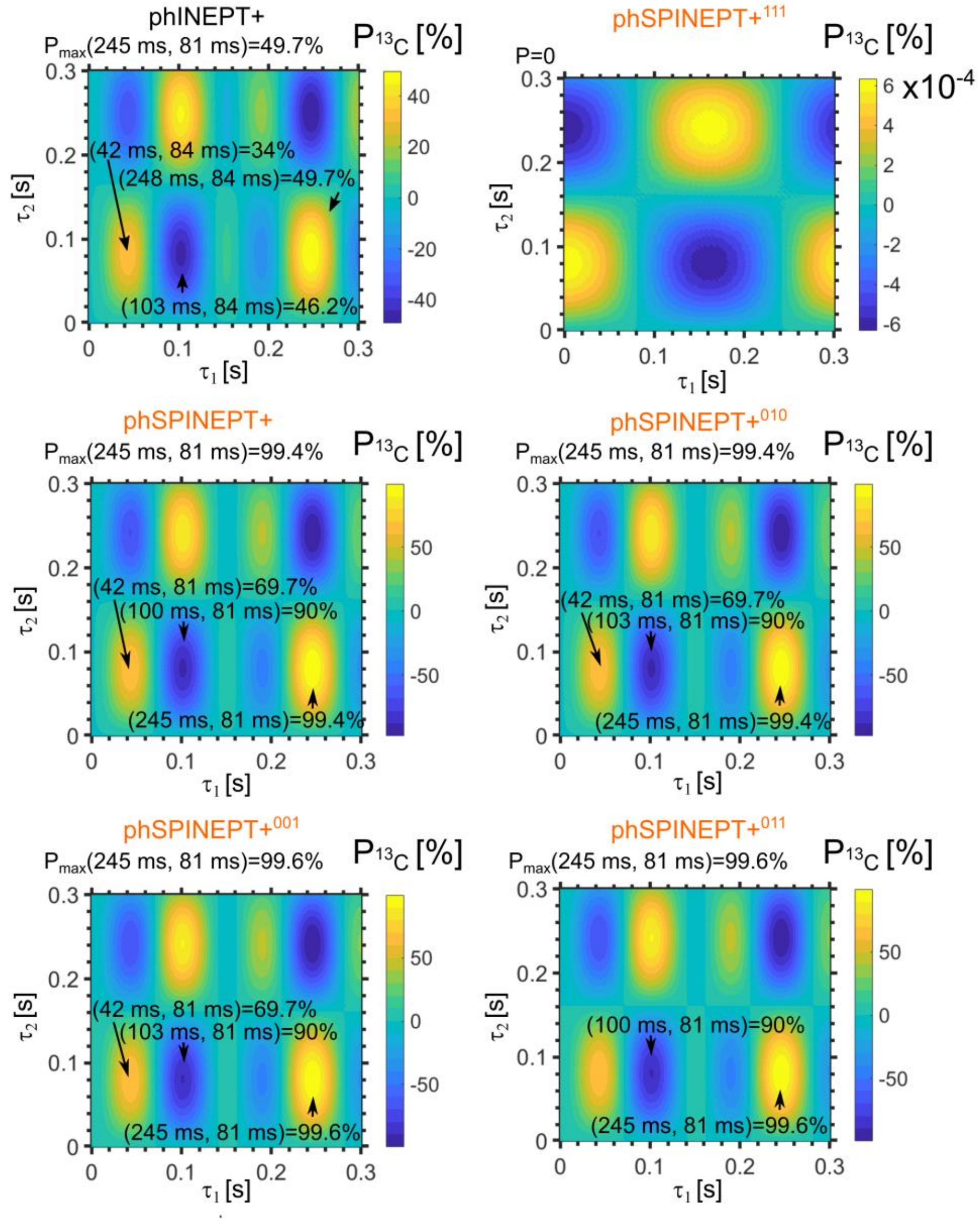


**Figure S6. $^{13}$C-polarization yield of 1-$^{13}$C-EA-$d_6$ using phINEPT+ or phSPINEPT+ as a function of the two evolution time intervals $\tau_1$ and $\tau_2$.** Two protons (stemming from $pH_2$) and one carbon at the acetate C1 position were considered (**Fig. S1**). The selective radio frequency pulses were set in resonance with the CHD protons of EA-$d_6$ at the chemical shift of $\delta$ = 4.1 ppm.

## D. 1-$^{13}$C-Vinyl pyruvate to 1-$^{13}$C-ethyl pyruvate

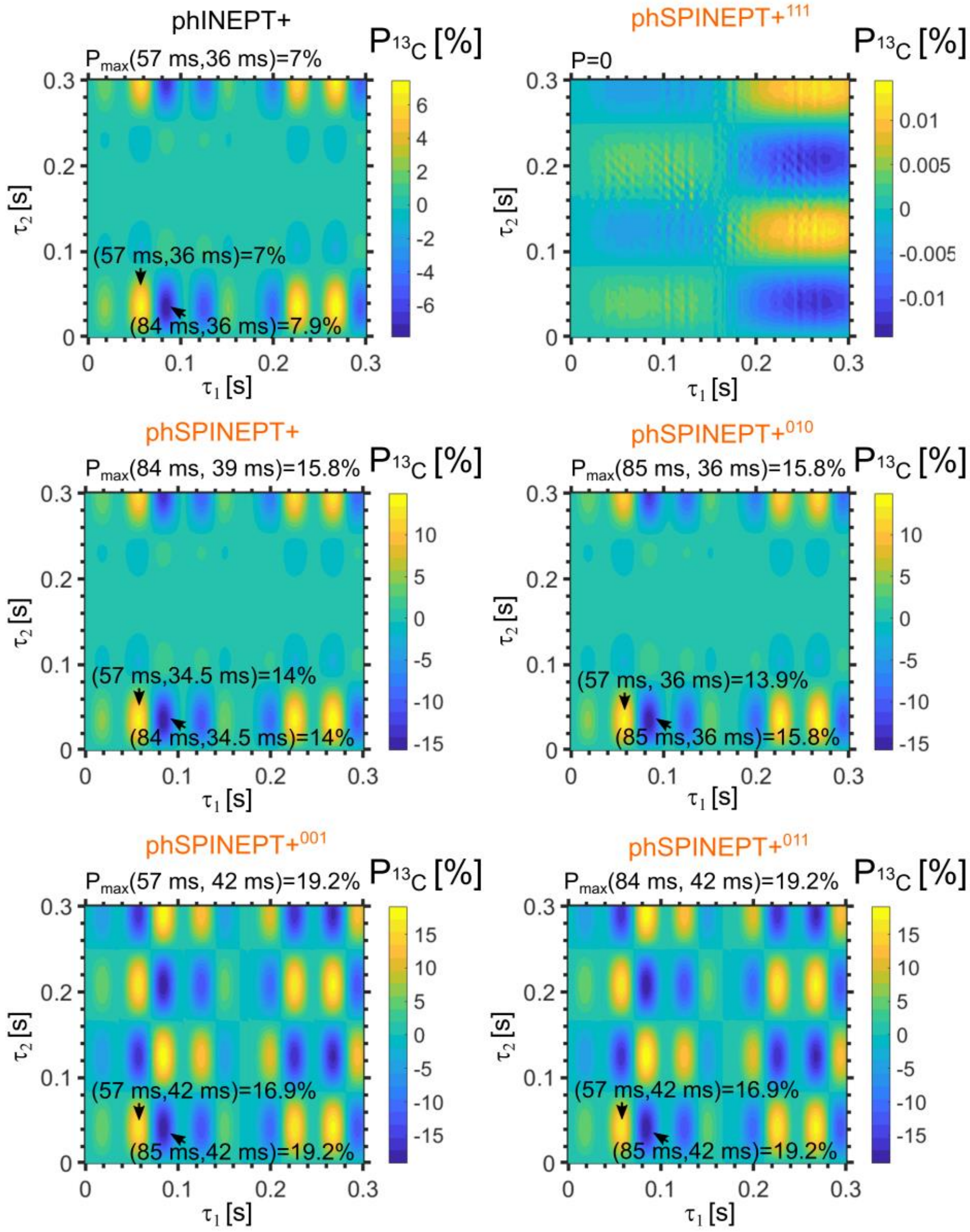


**Figure S7. $^{13}$C-polarization yield of 1-$^{13}$C-EP using phINEPT+ or phSPINEPT+ as a function of the two evolution time intervals $\tau_1$ and $\tau_2$.** Eight protons (two stemming from pH2) and one carbon at the acetate C1 position were considered (**Fig. S1**). The selective radio frequency pulses were set in resonance with the $CH_2$ protons of EP at the chemical shift of $\delta$ = 4.33 ppm.

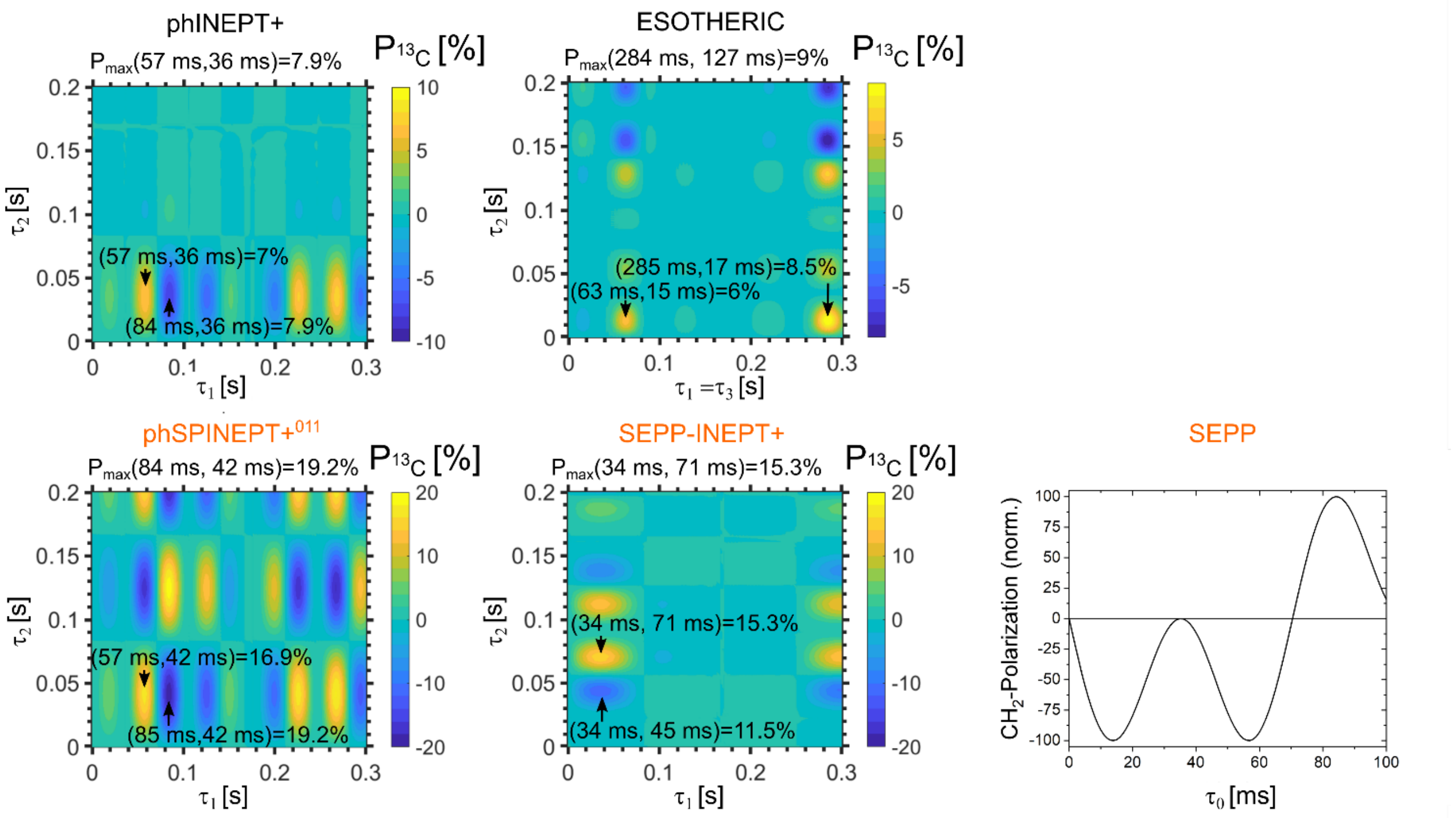


**Figure S8. $^{13}$C-polarization yield of 1-$^{13}$C-EP using phINEPT+, ESOTHERIC,[3] SEPP-INEPT+ and phSPINEPT+$^{011}$ as a function of the two evolution time intervals $\tau_1$ and $\tau_2$.** Eight protons (two stemming from $pH_2$) and one carbon at the acetate C1 position were considered (**Fig. S1**). The selective radio frequency pulses were set in resonance with the $CH_2$ protons of EP at the chemical shift of $\delta = 4.33$ ppm. Note that there are 6 evolution intervals in ESOTHERIC (instead of four in phINEPT+) sequence, from which we kept $\tau_1 = \tau_3$ as suggested by Korchak and coworkers; also, note that here $\tau$-intervals are half of the $\Delta$-intervals given in the ESOTHERIC scheme in the original work (see our Fig. S4).[3] The SEPP time interval $\tau_0$ was optimized to give the maximum net polarization of the $CH_2$ protons (SEPP plot) after the SEPP block and was set to $\tau_0 \cong 14$ ms for SEPP-INEPT+. Note that this value deviates from $\frac{1}{4J_{HH}} \cong 35$ ms.

**Table S2. Optimal SOT parameters $\tau_x$, total time of SOT $t_{total}$ and simulated polarization $P^{th}$ under this conditions for 1-$^{13}$C-EP**. Note that SEPP-INEPT is 16ms shorter than the phSPINEPT+$^{011}$ but provides 4 % less polarization.

| SOT sequence | $\tau_1$ [ms] | $\tau_2$ [ms] | $\tau_0$ / $\tau_3$ [ms] | $t_{total}$ [ms] | $P^{th}$ [%] |
|---|---|---|---|---|---|
| phINEPT+ | 84 | 36 | - | 240 | 8 |
| phSPINEPT+$^{011}$ | 85 | 42 | - | 254 | 19.2 |
| ESOTHERIC | 285 | 17 | 285 | 1212 | 8.5 |
| SEPP-INEPT+ | 34 | 71 | 14 | 238 | 15.3 |

## E. 1-$^{13}$C-Vinyl pyruvate-d$_6$ to 1-$^{13}$C-ethyl pyruvate-d$_6$

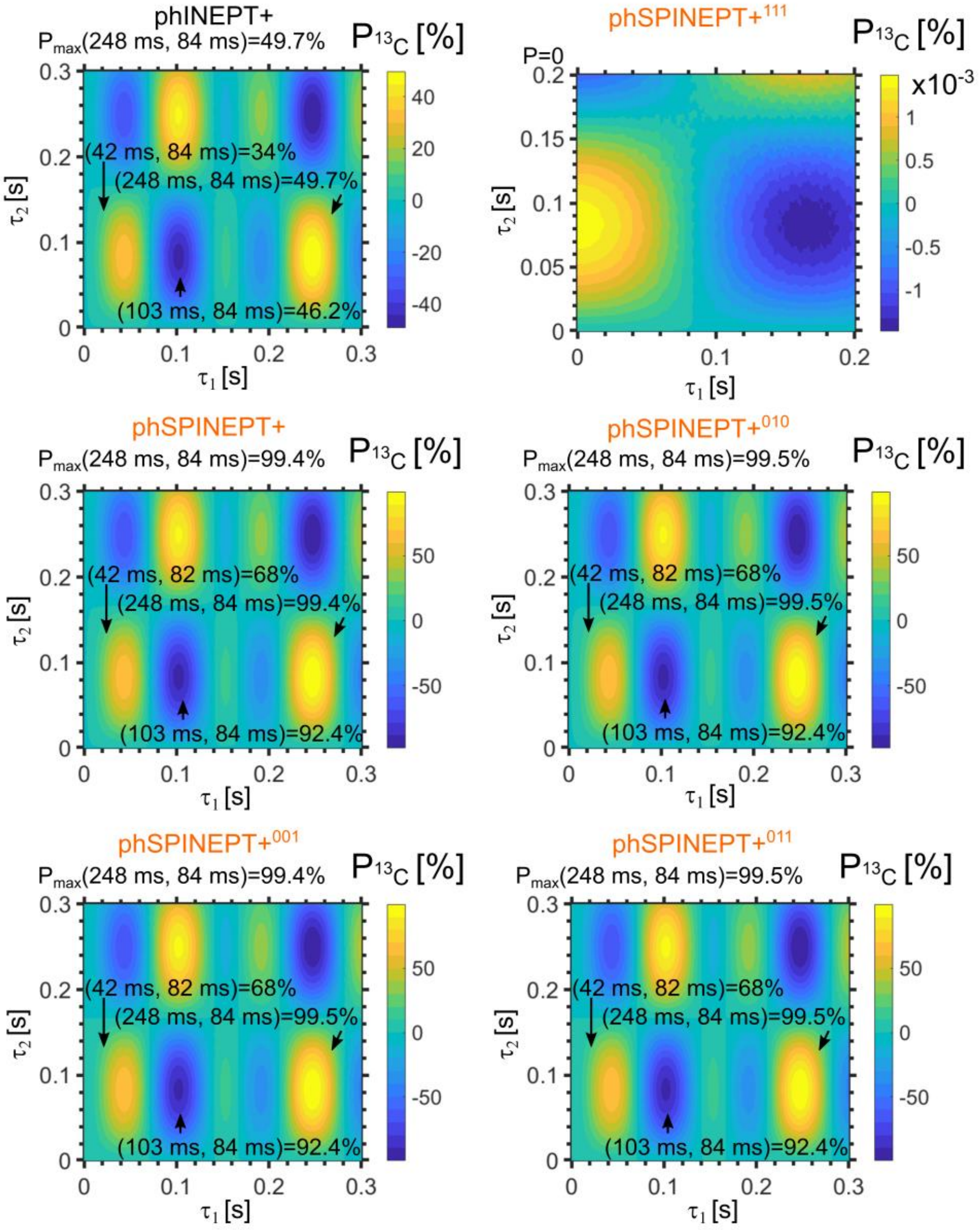


**Figure S9. $^{13}$C-polarization yield of 1-$^{13}$C-EP-d$_6$ using phINEPT+ or phSPINEPT+ as a function of the two evolution time intervals $\tau_1$ and $\tau_2$.** Two protons (two stemming from $pH_2$) and one carbon at the acetate C1 position were considered (**Fig. S1**). The selective radio frequency pulses were set in resonance with the CHD protons of EP-d$_6$ at the chemical shift of $\delta$ = 4.33 ppm.

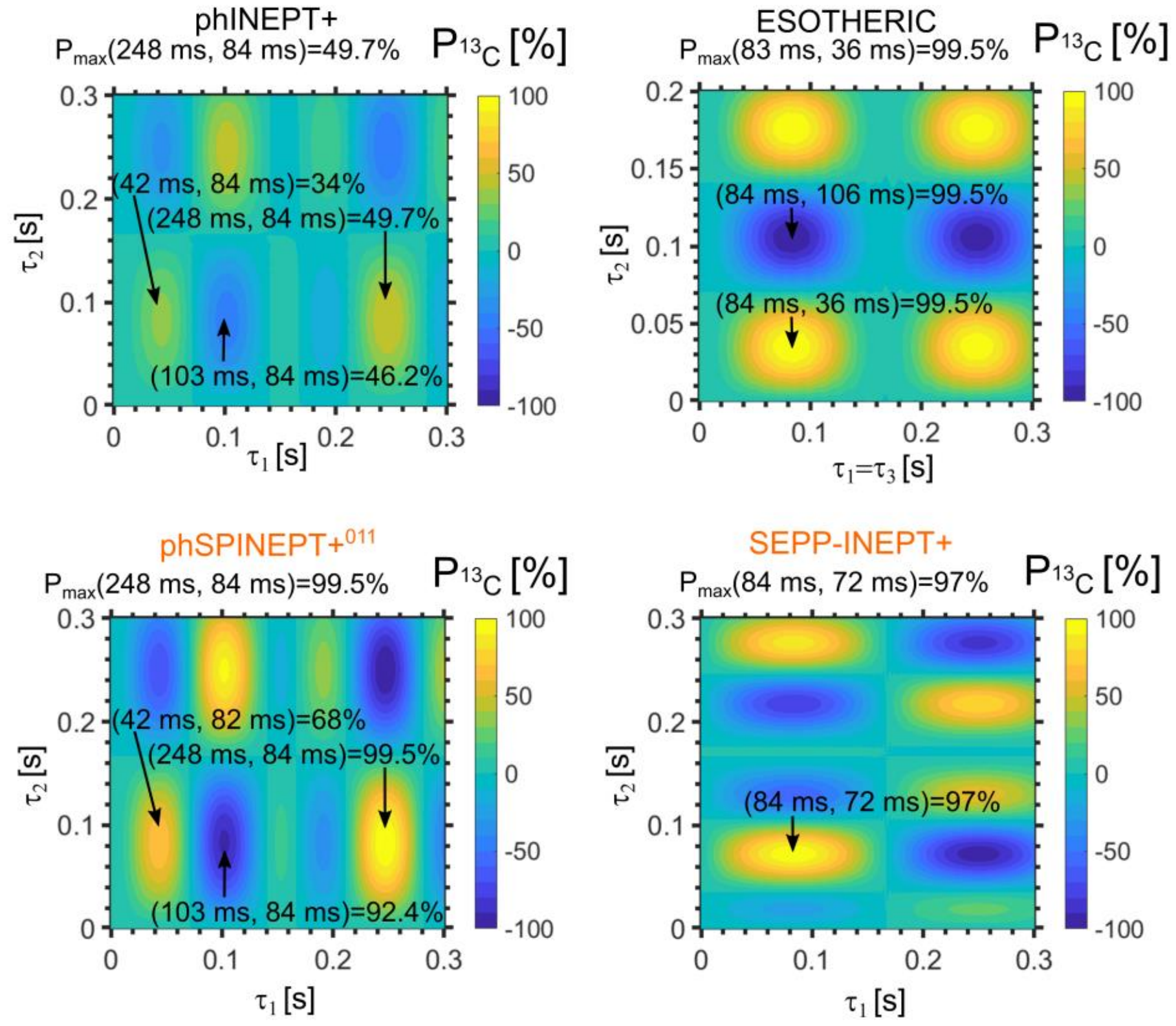


**Figure S10. $^{13}$C-polarization yield of 1-$^{13}$C-EP-d$_6$ phINEPT+, ESOTHERIC,[3] SEPP-INEPT+ and phSPINEPT+$^{011}$ as a function of the two evolution time intervals $\boldsymbol{\tau_1}$ and $\boldsymbol{\tau_2}$.** Two protons (two stemming from p$H_2$) and one carbon at the acetate C1 position were considered (**Fig. S1**). The selective radio frequency pulses were set in resonance with the CHD protons of EP-d$_6$ at the chemical shift of $\delta$ = 4.33 ppm. Note that there are 6 evolution intervals in ESOTHERIC (instead of four in phINEPT+) sequence, from which we kept $\boldsymbol{\tau_1} = \boldsymbol{\tau_3}$ as suggested by Korchak and coworkers; also, note that here $\boldsymbol{\tau}$-intervals are half of the $\boldsymbol{\Delta}$-intervals given in the ESOTHERIC scheme in the original work (see our fig. S4).[3] SEPP interval for SEPP-INEPT+ was $\boldsymbol{\tau_0} = \frac{\mathbf{1}}{\mathbf{4}\boldsymbol{J_{HH}}}$ ≅35 ms.

**Table S3. Optimal SOT parameters $\boldsymbol{\tau_x}$, total time of SOT $\boldsymbol{t_{total}}$ and simulated polarization $\boldsymbol{P^{th}}$ under this conditions for 1-$^{13}$C-EP-d$_6$**. Note that SEPP-INEPT is 34 ms shorter than the phSPINEPT+$^{011}$ and provides 5 % more polarization. A polarization of ~100% is reached with phSPINEPT+$^{011}$ with a longer sequence ($t_{total}$ = 664 ms).

| SOT sequence | $\boldsymbol{\tau_1}$ [ms] | $\boldsymbol{\tau_2}$ [ms] | $\boldsymbol{\tau_0}$ / $\boldsymbol{\tau_3}$ [ms] | $\boldsymbol{t_{total}}$ [ms] | $\boldsymbol{P^{th}}$ [%] |
|---|---|---|---|---|---|
| phINEPT+ | 103 | 84 | - | 374 | 46.2 |
| phSPINEPT+$^{011}$ | 103 | 84 | - | 374 | 92 |
| ESOTHERIC | 84 | 36 | 84 | 408 | 100 |
| SEPP-INEPT+ | 84 | 72 | 14 | 340 | 97 |

## F. 1-$^{13}$C- Hydroxyethyl acrylate-d$^{8}$ to 1-$^{13}$C- Hydroxyethyl propionate-d$_8$

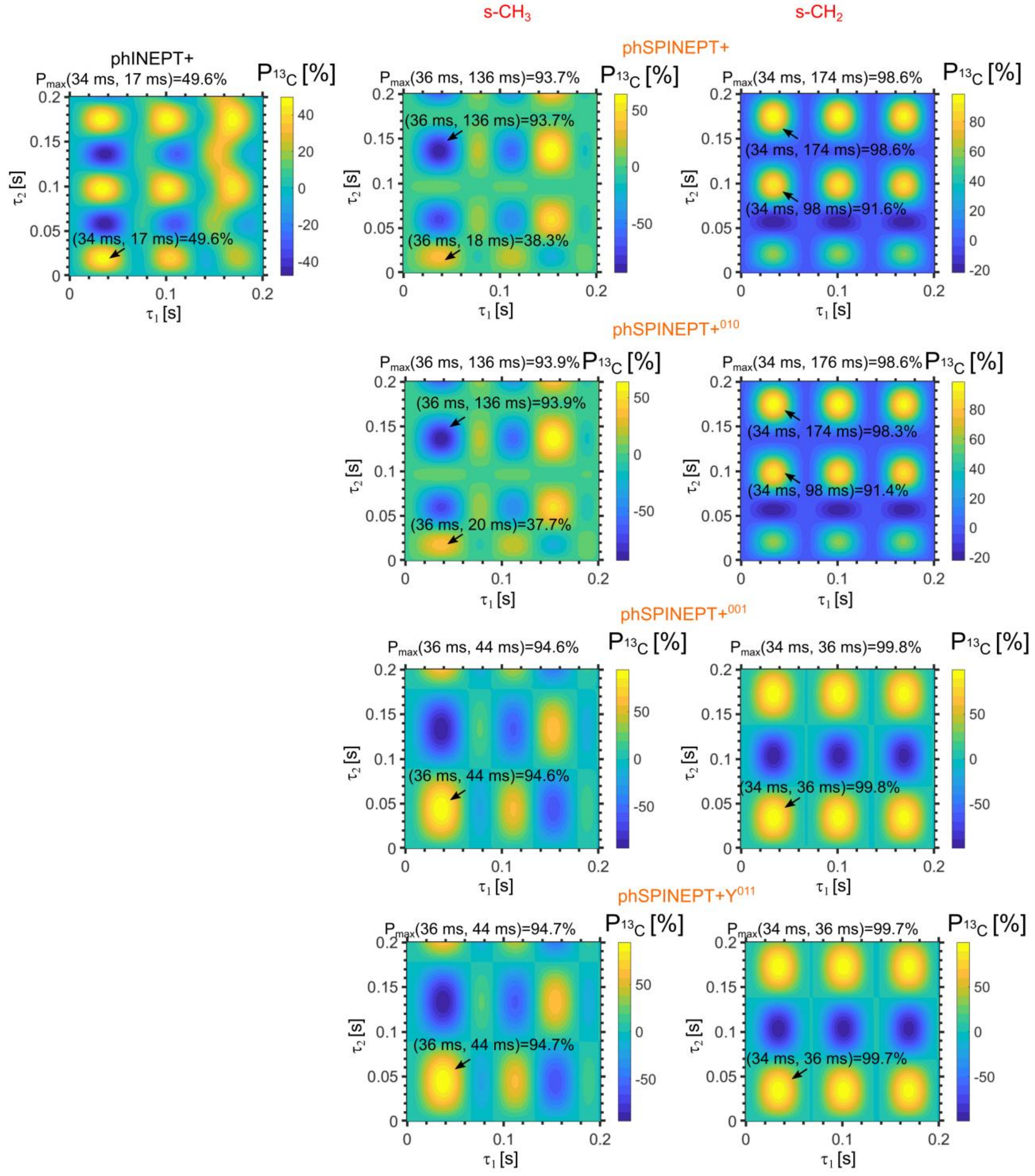


**Figure S11. $^{13}$C-polarization yield of 1-$^{13}$C-HEP-d$_8$ using phINEPT+ or phSPINEPT+ as a function of the two evolution time intervals $\tau_1$ and $\tau_2$.** Two ethyl protons (stemming from pH$_2$) and one carbon at C1 position were considered (**Fig. S1**). The selective radio frequency pulses were set in resonance with the CH$_2$ (s-CH$_2$; at the chemical shift $\delta$ = 2.35 ppm) or CH$_3$ (s-CH$_3$; at $\delta$ = 1.15 ppm) protons of 1-$^{13}$C-HEP. Note that here complete deuteration of HEA was assumed although experimentally only 1-$^{13}$C-HEA-d$_3$ was available (i.e. the HOCH$_2$CH$_2$O- fragment was not deuterium labeled).

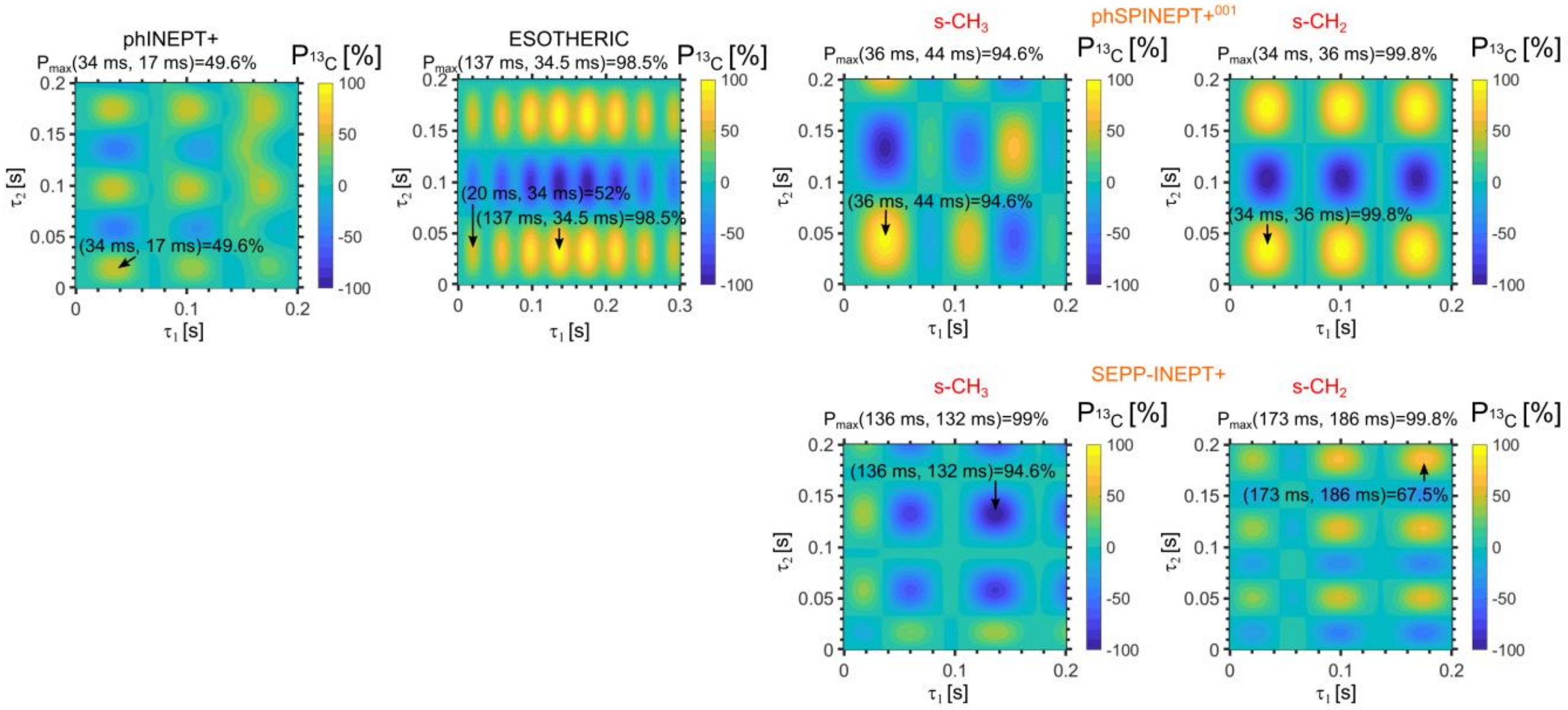


**Figure S12. $^{13}$C-polarization yield of 1-$^{13}$C-HEP-d$_8$ phINEPT+, ESOTHERIC,[3] SEPP-INEPT+ and spINEPT+$^{011}$ as a function of the two evolution time intervals $\tau_1$ and $\tau_2$.** Two protons (stemming from pH$_2$) and one carbon at C1 position were considered (Fig. S1). The selective radio frequency pulses were set in resonance with the CH2 (s-CH$_2$; at the chemical shift $\delta = 2.35$ ppm) or CH$_3$ (s-CH$_3$; at $\delta = 1.15$ ppm) protons of 1-$^{13}$C-HEP. Note that here complete deuteration of HEA was assumed although experimentally only 1-$^{13}$C-HEA-d3 was available available (i.e. the HOCH$_2$CH$_2$O- fragment was not deuterium labeled). Note that there are 6 evolution intervals in ESOTHERIC (instead of four in phINEPT+) sequence, from which we kept $\tau_1 = \tau_3$ as suggested by Korchak and coworkers; also, note that here $\tau$-intervals are half of the $\Delta$-intervals given in the ESOTHERIC scheme in the original work (see our Fig. S4).[3] SEPP interval for SEPP-INEPT+ was $\tau_0 = \frac{1}{4J_{HH}} \cong 33$ ms.

**Table S4. Optimal SOT parameters $\tau_x$, total time of SOT $t_{total}$ and simulated polarization $P^{th}$ under this conditions for 1-$^{13}$C-HEP-d$_8$.** Note that here ESOTHERIC and SEPP-INEPT+ are much longer than phSPINEPT+$^{011}$, which provides polarization close to unity. phINEPT+ is shorter but provides only half the polarization. This result and Fig. S14 show that phSPINEPT+$^{011}$ is effective on both, molecules with a "symmetric" and "asymmetric" *J*-coupling pattern.

| SOT sequence | $\tau_1$ [ms] | $\tau_2$ [ms] | $\tau_0$ / $\tau_3$ [ms] | $t_{total}$ [ms] | $P^{th}$ [%] |
|---|---|---|---|---|---|
| phINEPT+ | 34 | 17 | - | 96 | 50 |
| phSPINEPT+$^{011}$ | 34 | 36 | - | 140 | 99 |
| ESOTHERIC | 137 | 34.5 | 137 | 617 | 99 |
| SEPP-INEPT+ | 136 | 132 | 33 | 602 | 95 |

## G. 1-$^{13}$C- Hydroxyethyl acrylate-$d_3$ to 1-$^{13}$C- Hydroxyethyl propionate-$d_3$

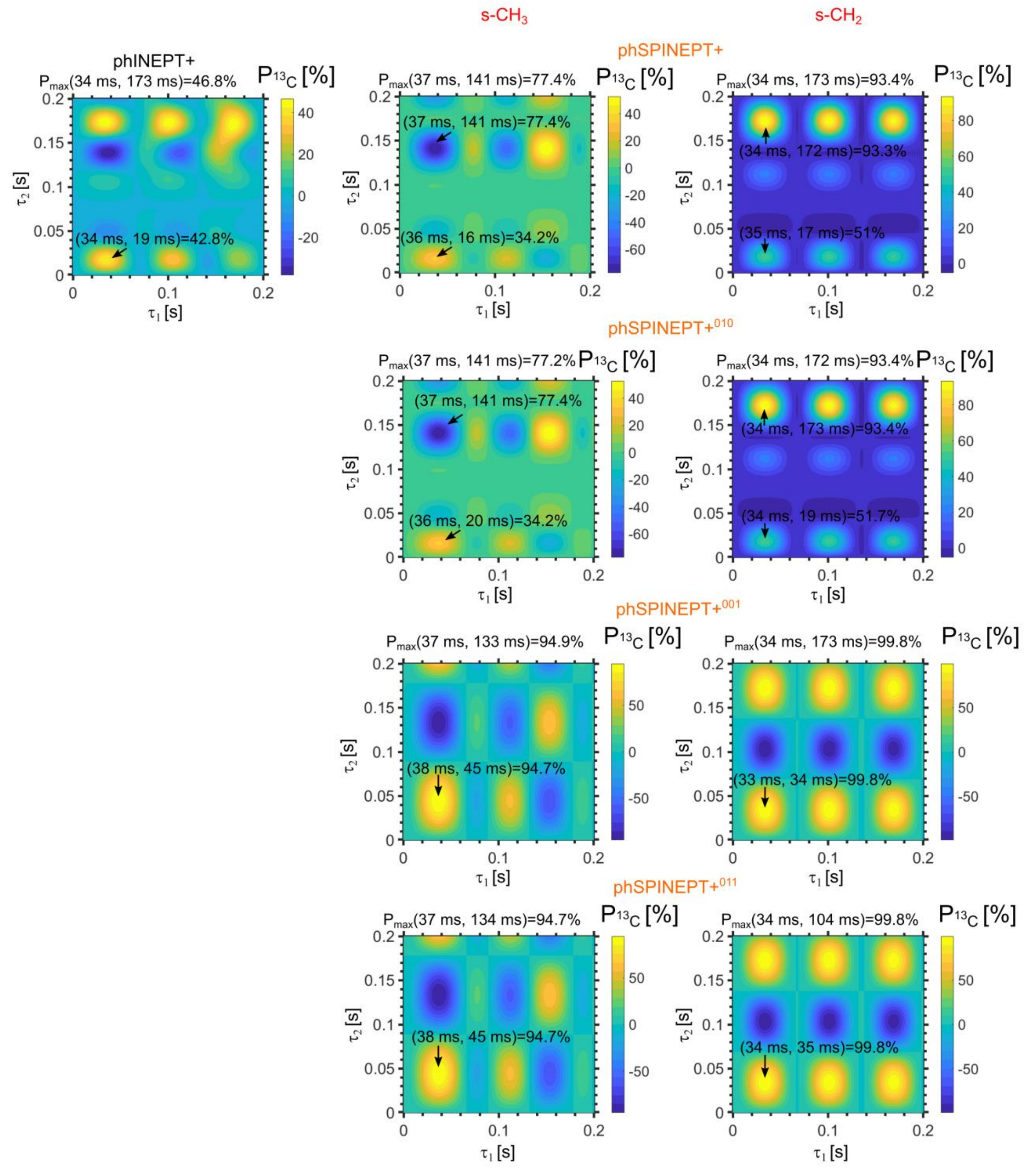


**Figure S13. $^{13}$C-polarization yield of 1-$^{13}$C-HEP-$d_3$ using phINEPT+ or phSPINEPT+ as a function of the two evolution time intervals $\boldsymbol{\tau_1}$ and $\boldsymbol{\tau_2}$.** Two ethyl protons (stemming from $pH_2$), one carbon at C1 position and two other nearest methylene protons were considered (**Fig. S1**). The selective radio frequency pulses were set in resonance with the $CH_2$ (s-$CH_2$; at the chemical shift $\delta$ = 2.35 ppm) or $CH_3$ (s-$CH_3$; at $\delta$ = 1.15 ppm) protons of 1-$^{13}$C-HEP. Note that two other methylene protons were neglected because they do not have observable coupling with C1.

## H. Efficiency of phSPINEPT+ in various three spin systems

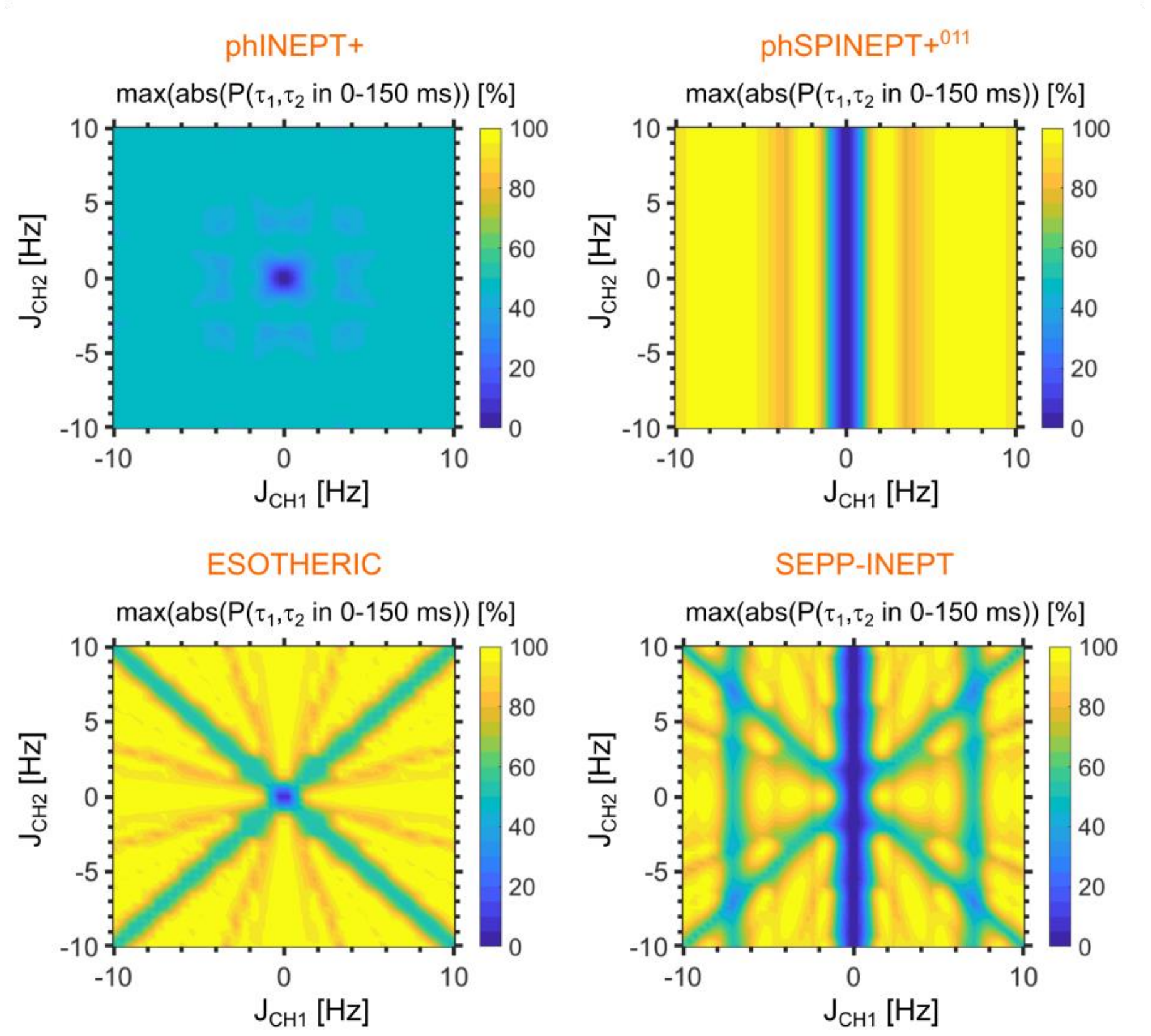


**Figure S14. $^{13}C$-polarization yield using phSPINEPT+$^{011}$ in arbitrary three spin system as a function of the H-C spin-spin coupling constants.** For each pair of $J_{CH1}$ and $J_{CH2}$ the phSPINEPT+$^{011}$ polarization was calculated as a function of the evolution timings $\boldsymbol{\tau_1}$ and $\boldsymbol{\tau_2}$. From this data, the absolute maximum of the polarization was extracted and is displayed in this figure. $\boldsymbol{\tau_1}$ and $\boldsymbol{\tau_2}$ intervals were varied in the range of 0 to 150 ms in steps of 2 ms. Proton chemical shift difference was 3 ppm, their mutual spin-spin coupling, $J_{H1H2}$, was fixed at 7 Hz, magnetic field was set to $\boldsymbol{B_0}$ = 9.4 T. The selective pulses were applied on the H1 nucleus.

### 4.3. phINEPT+ and phSPINEPT+ spectra (Experimental)

#### A. 1-$^{13}$C-Vinyl acetate to 1-$^{13}$C-ethyl acetate

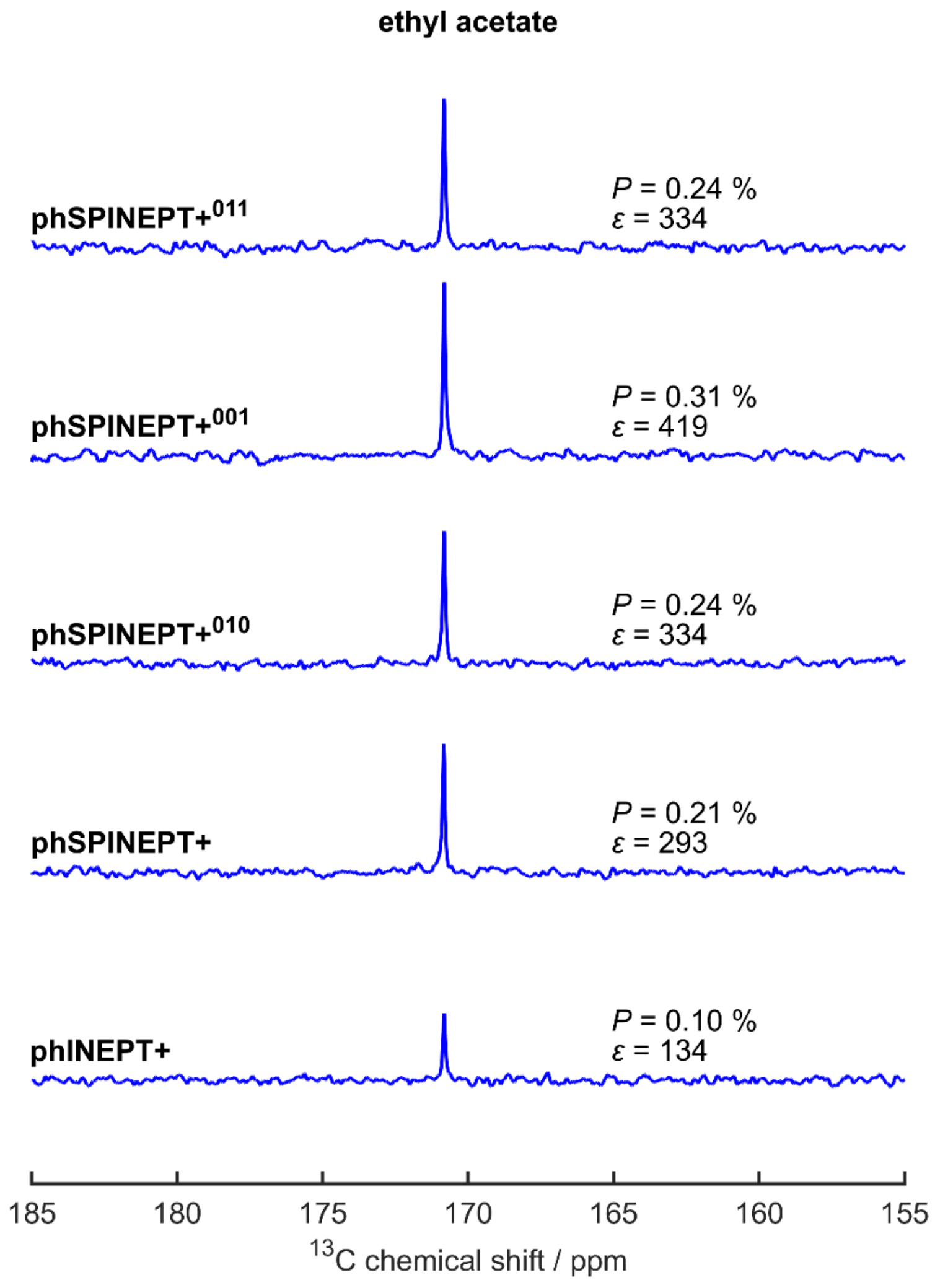


**Figure S15. $^{13}$C-NMR spectra of 1-$^{13}$C-hyperpolarized ethyl acetate using the phINEPT+ and phSPINEPT+ SOT sequences.** The used experimental parameters are summarized in **Tab S1**.

## B. 1-$^{13}$C-Vinyl acetate-d$_6$ to 1-$^{13}$C-ethyl acetate-d$_6$

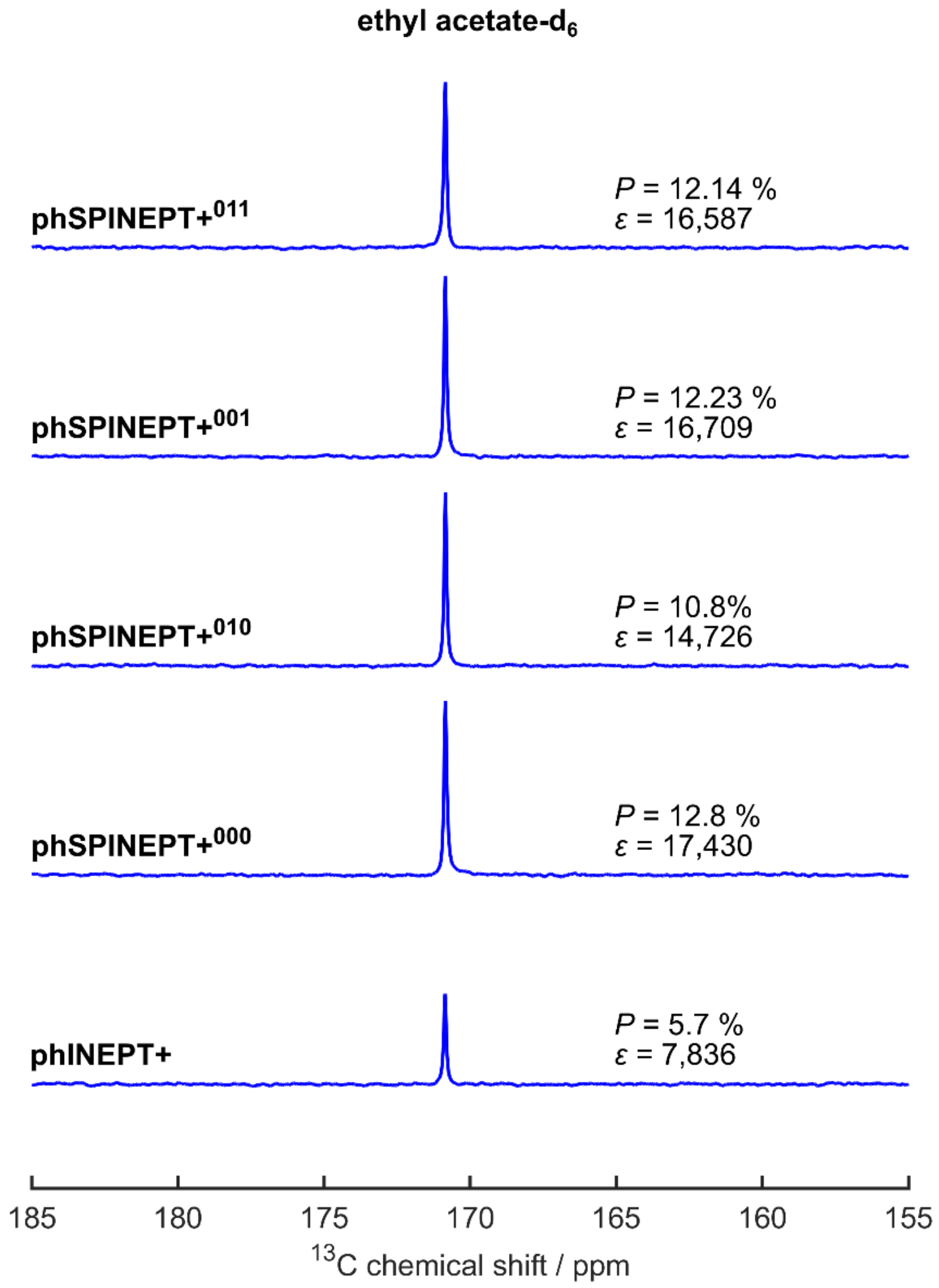


**Figure S16. $^{13}$C-NMR spectra of 1-$^{13}$C-hyperpolarized ethyl acetate-d$_6$ using the phINEPT+ and phSPINEPT+ SOT sequences.** The used experimental parameters are summarized in **Tab S1**.

### C. 1-$^{13}$C-Vinyl pyruvate to 1-$^{13}$C-ethyl pyruvate

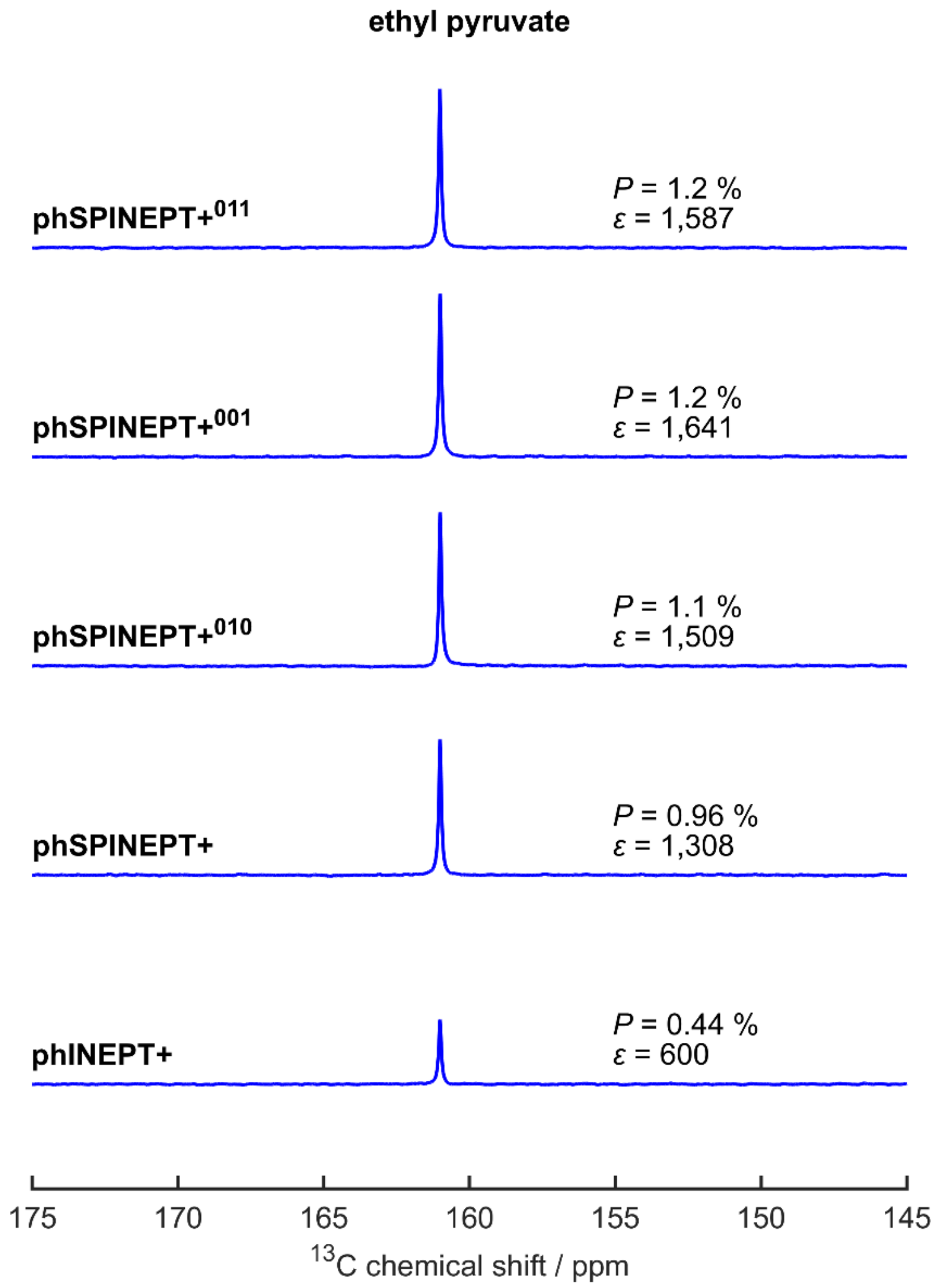


**Figure S17 $^{13}$C-NMR spectra of 1-$^{13}$C-hyperpolarized ethyl pyruvate using the phINEPT+ and phSPINEPT+ SOT sequences.** The used experimental parameters are summarized in **Tab S1**.

## D. 1-$^{13}$C-hydroxyethyl acrylate-$d_3$ to 1-$^{13}$C-hydroxyethyl propionate-$d_3$

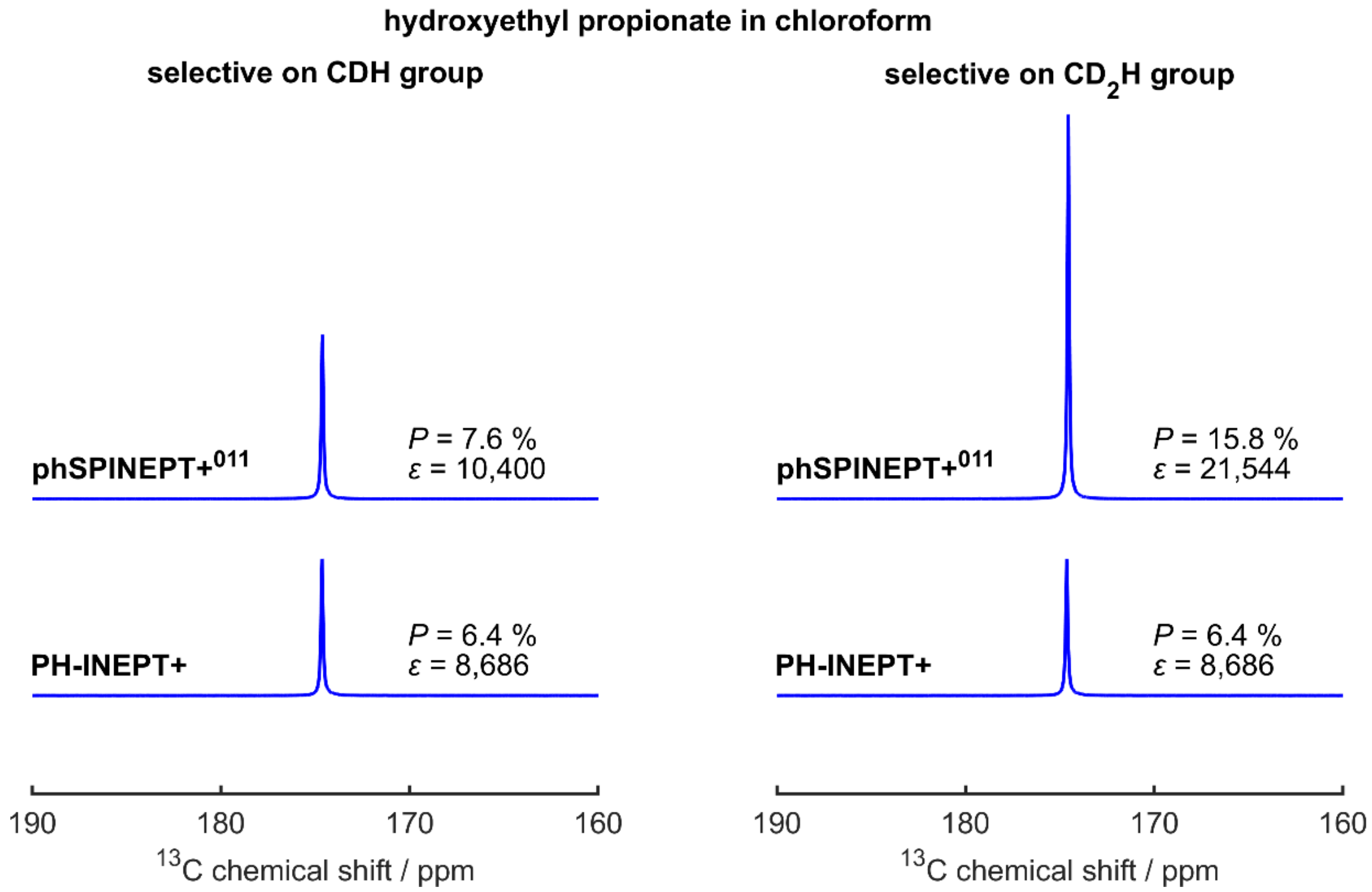


**Figure S18. $^{13}$C-NMR spectra of 1-$^{13}$C-hyperpolarized hydroxyethyl propionate-$d_3$ in chloroform-d using the phINEPT+ and phSPINEPT+ SOT sequences.** The used experimental parameters are summarized in **Tab S1**.

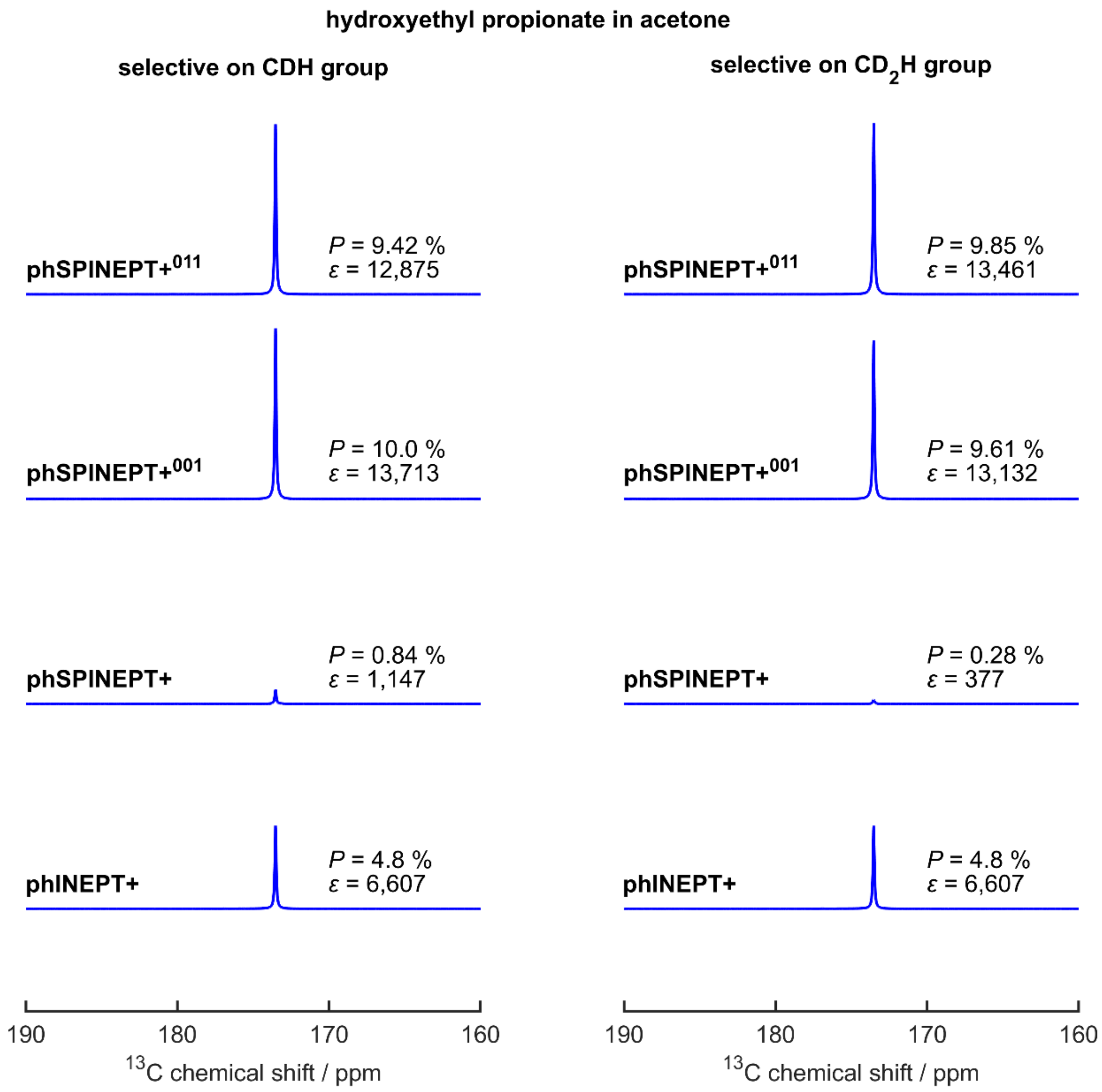


**Figure S19. $^{13}$C-NMR spectra of 1-$^{13}$C-hyperpolarized hydroxyethyl propionate-$d_3$ in acetone-$d_6$ using the phINEPT+ and phSPINEPT+ SOT sequences.** Not all experimentally used parameters were identical to the one that according to our simulations give optimal polarization. The used experimental parameters are summarized in **Tab S1** and correpsonding polarization maps are presented in **Fig S12**.

## 5. Estimation of hydrogenation reaction and magnetization decay

We consider a reaction of substrate $A$ to product $B$ which creates magnetization $M_B$ that subsequently relaxes to 0 with rate $R$:

$A \xrightarrow{k_{hyd}} B$ [eq S1]

$M_B \xrightarrow{R} 0$ [eq S2]

The concentration of product at a time t is

$[B](t) = [A_0](1 - e^{-tk_{hyd}})$ [eq S3]

The change of concentration of product B in the interval (t,t+dt) is

$d[B](t) = [A_0]k_{hyd}e^{-tk_{hyd}}dt$ [eq S4]

Then magnetization (without nuclear dipol moment for simplicity or in the units of polarization times concentration) that is generated in the same time interval is given by

$dM_B(t,t) = P_{max}d[B](t) = [A_0]k_{hyd}e^{-tk_{hyd}}dt$ [eq S5]

The remaining magnetization at time point $\tau$ which was generated in the interval (t,t+dt) is

$dM_B(t,\tau) = [A_0]P_{max}k_{hyd}e^{-R(\tau-t)}e^{-tk_{hyd}}dt$ [eq S6]

The total magnetization at time point $\tau$ then can be calculated by a simple integration over t:

$$M_B(\tau) = P(\tau)[B](\tau) = \int_0^\tau dM_B(t,\tau)dt = [A_0]P_{max}k_{hyd}e^{-R\tau}\int_0^\tau e^{-t(k_{hyd}-R)}dt =$$

$= [A_0]P_{max}\frac{1}{1-R/k_{hyd}}\left(e^{-R\tau} - e^{-\tau k_{hyd}}\right)$ [eq S7]

And in terms of remaining polarization that is usually reported in the literature

$P(\tau) = P_{max}\frac{1}{1-\frac{R}{k_{hyd}}}\frac{e^{-R\tau}-e^{-\tau k_{hyd}}}{1-e^{-\tau k_{hyd}}}$ [eq S8]

If we assume that $R \ll k_{hyd}$, which is justified because $R \sim \frac{1}{T_1} < 0.25\ \mathrm{s}^{-1}$ for all measured $T_1$ values (Fig. S.1) and in ≈6 seconds we achieved complete hydrogenation (i.e. $k_{\mathrm{hyd}} \geq \frac{5}{6}\mathrm{s}^{-1} \approx 0.8\ \mathrm{s}^{-1}$), then

$[B](\tau \gg 1/k_{hyd}) \cong [A_0]$ [eq S9]

And

$M_B\left(\frac{1}{R} \gg \tau \gg \frac{1}{k_{hyd}}\right) = P(\tau)[B](\tau) \cong [A_0]P_{max}e^{-R\tau}$ [eq S10]

Then polarization is

$P(\tau) \cong P_{max}e^{-R\tau}$ [eq S11]

Under the assumption of a very fast hydrogenation compared to relaxation equations eq S7 and S8 simplify to eq S10 and S11 respectively. Eq S11 was used in the main manuscript.

To make sure that in the model no division by zero happens, let us also consider the case when $k_{hyd} = R + \Delta$ with $\Delta \ll R, k_{hyd}$, i.e. when $k_{hyd} \cong R$.

Then simplification of eq S7 results in

$$M_B\left(\tau, k_{hyd} \cong R\right) = [A_0]P_{max}\frac{e^{-R\tau}}{1-\frac{1}{1+\frac{\Delta}{R}}}(1 - e^{-\tau\Delta}) = [A_0]P_{max}\frac{e^{-R\tau}}{1-1+\frac{\Delta}{R}}(1 - 1 + \tau\Delta) =$$

$$= [A_0]P_{max}\tau R e^{-R\tau} \quad \text{[eq S12]}$$

The $M_B\left(\tau, k_{hyd} \cong R\right)$ reaches its maximum of $[A_0]P_{max}e^{-1}$ when $\tau = 1/R$.

Under the same conditions eq S8 gives

$$P\left(\tau, k_{hyd} \cong R\right) = P_{max}\frac{e^{-R\tau}}{1-\frac{1}{1+\Delta/R}}\frac{1-e^{-\tau\Delta}}{1-e^{-\tau R}} = P_{max}\frac{e^{-R\tau}}{1-1+\frac{\Delta}{R}}\frac{1-1+\tau\Delta}{1-e^{-\tau R}} = P_{max}\frac{R\tau}{e^{R\tau}-1} \quad \text{[eq S13].}$$

The function reaches maximum (for positive $\tau$) at $\tau = 0$. Hence to maximise polarization this conditions must be avoided and $k_{hyd}$ should be much faster than R.

## 6. Scripts: pulse sequences (Bruker code)

Below, the general SOT scheme with $^1$H decoupling is given, and all variants of used SOTs.

### A. SOT frame

```
#include <Avance.incl>
"acqt0=-p1*2/3.1416"
"p11=p1/2" ; 13C:45
"p12=p1*2" ; 13C:180
"p21=p2/2" ; 1H: 45
"p22=p2*2" ; 1H: 180
1 ze
2 30m do:f2
        d1
        10m LOCKH_ON
        30m pl1:f1
        30m pl2:f2
        ; put your pH2 bubbling script here
        ; <SOT script>
        go=2 ph31
        30m LOCKH_OFF do:f2 mc #0 to 2 F0(zd)
exit
ph1=0
ph2=1
ph31=0
```

### B. SOT: phINEPT+

```
(p21 ph2):f2
        d5 ; d5 = tau1/2
(center (p12 ph1):f1 (p22 ph1):f2)
        d5 ; d5 = tau1/2
(center (p1 ph1):f1  (p2 ph2):f2)
        d6 ; d6 = tau2/2
        1u
(center (p12 ph1):f1            (p22 ph1):f2)
        d6 pl12:f2 ; d6 = tau2/2
1u cpd2:f2
```

### C. SOT: phSPINEPT+

```
(p10:sp1 ph2):f3
        d4 pl2:f3 ; d4 = tau1/2-p10/2
(center (p12 ph1):f1 (p22 ph1):f3)
        d5 ; d5 = tau1/2
(center (p1 ph1):f1  (p2 ph2):f3)
        d6 ; d6 = tau2/2
        1u
(center (p12 ph1):f1 (p22 ph1):f3)
        d6 pl12:f2 ; d6 = tau2/2
1u cpd2:f2
```

### D. SOT: phSPINEPT+[001]

```
(p10:sp1 ph2):f3
        d4 pl2:f3 ; d4 = tau1/2-p10/2
(center (p12 ph1):f1 (p22 ph1):f3)
        d5 ; d5 = tau1/2
(center (p1 ph1):f1 (p2 ph2):f3)
        d6 ; d6 = tau2/2-p10/2
        1u
(center (p12 ph1):f1 (p10:sp2 ph1):f3)
        d7 pl12:f2 ; d7 = tau2/2-p10/2
1u cpd2:f2
```

### E. phSPINEPT+[010]

```
(p10:sp1 ph2):f3
        d4 pl2:f3 ; d4 = tau1/2-p10/2
(center (p12 ph1):f1 (p22 ph1):f3)
        d5 ; d5 = tau1/2-p10/2
(center (p1 ph1):f1 (p10:sp1 ph2):f3)
        d6 pl2:f3 ; d6 = tau2/2-p10/2
        1u
```

```
(center (p12 ph1):f1 (p22 ph1):f3)
        d7 pl12:f2 ; d7 = tau2/2
1u cpd2:f2
```

### F. phSPINEPT+[011]

```
(p10:sp1 ph2):f3
        d4 pl2:f3 ; d4 = tau1/2-p10/2
(center (p12 ph1):f1 (p22 ph1):f3)
        d5 ; d5 = tau1/2-p10/2
(center (p1 ph1):f1 (p10:sp1 ph2):f3)
        d6 ; d6 = tau2/2-p10
        1u
(center (p12 ph1):f1 (p10:sp2 ph1):f3)
        d7 pl12:f2 ; d7 = tau2/2
1u cpd2:f2
```